\documentclass[12pt, onecolumn]{IEEEtran}%[journal]{IEEEtran}%
\usepackage{epsfig,latexsym,amssymb,amsmath,graphics,color}
\usepackage{subfigure}
\usepackage{float}
\usepackage{verbatim}
\usepackage{multicol}
\usepackage{amstext}
\usepackage{mathrsfs}
\usepackage[all]{xy}
\usepackage{cite}
\usepackage{setspace}
\usepackage{amsthm}
\usepackage{graphicx}

\usepackage[ruled]{algorithm2e}
\usepackage{algpseudocode}
\usepackage{booktabs}
\usepackage[sans]{dsfont}
\usepackage{txfonts}
\usepackage{amsfonts}
\usepackage{upgreek}
\usepackage{multirow,multicol,array}
\usepackage{geometry}

\newcommand{\tabincell}[2]{\begin{tabular}{@{}#1@{}}#2\end{tabular}}
\newtheorem{prop}{Proposition}

\newtheorem{cor}{Corollary}

\newtheorem{lm}{Lemma}

\newtheorem{thm}{Theorem}

\newcommand{\be}{\begin{eqnarray}}
\newcommand{\ee}{\end{eqnarray}}
\newcommand{\benn}{\begin{eqnarray*}}
\newcommand{\eenn}{\end{eqnarray*}}
\def\IR{\rm I \kern-0.20em R}
\newcommand{\utwi}[1]{\mbox{\boldmath $ #1$}}

\newcommand{\bthm}{\begin{thm}}
\newcommand{\ethm}{\end{thm}}

\newcommand{\bcor}{\begin{cor}}
\newcommand{\ecor}{\end{cor}}
\newcommand{\bprop}{\begin{prop}}
\newcommand{\eprop}{\end{prop}}
\newcommand{\blm}{\begin{lm}}
\newcommand{\elm}{\end{lm}}
\newcommand{\beq}{\begin{equation}}
\newcommand{\eeq}{\end{equation}}
\newcommand{\ber}{\begin{eqnarray}}
\newcommand{\eer}{\end{eqnarray}}

\newcommand{\bproof}{\begin{proof}}
\newcommand{\eproof}{\end{proof}}

\newcommand{\bit}{\begin{itemize}}
\newcommand{\eit}{\end{itemize}}
\newcommand{\ben}{\begin{enumerate}}
\newcommand{\een}{\end{enumerate}}
\newcommand{\bdesc}{\begin{description}}
\newcommand{\edesc}{\end{description}}
\newcommand{\beqarrn}{\begin{eqnarray*}}
\newcommand{\eeqarrn}{\end{eqnarray*}}
\newcommand{\bproofof}{\begin{proofof}}
\newcommand{\eproofof}{\end{proofof}}
\newenvironment{rem}{\begin{trivlist}\item[]{\bf
Remark:}\hspace{4mm}}{\end{trivlist}}
\newcommand{\brem}{\begin{rem}}
\newcommand{\erem}{\end{rem}}
\newenvironment{rems}{\begin{trivlist}\item[]{\bf
Remarks}\begin{itemize}}{\end{itemize}\end{trivlist}}
\newcommand{\brems}{\begin{rems}}
\newcommand{\erems}{\end{rems}}
\newtheorem{fact}{Fact}
\newcommand{\bfact}{\begin{fact}}
\newcommand{\efact}{\end{fact}}
\newtheorem{examp}{Example}
\newcommand{\bexamp}{\begin{examp}\rm}
\newcommand{\eexamp}{\end{examp}}
\newtheorem{defn}{Definition}
\newcommand{\bdefn}{\begin{defn}\rm}
\newcommand{\edefn}{\end{defn}}

\newtheorem{alg}{Algorithm}
\newcommand{\balg}{\begin{alg}}
\newcommand{\ealg}{\end{alg}}

\newtheorem{prob}{Problem}
\newcommand{\bprob}{\begin{prob}}
\newcommand{\eprob}{\end{prob}}

\newcommand{\bvtm}{\begin{verbatim}}
\newcommand{\bfig}{\begin{figure}}
\newcommand{\efig}{\end{figure}}
\newcommand{\bcen}{\begin{center}}
\newcommand{\ecen}{\end{center}}

\long\def\comment#1{}

\def \n2{{N_0 \over 2}}

\def \h5{\hspace{0.5in}}

\newcommand{\bS}{{\utwi{S}}}

\title{A Double-station Access Protocol for Optical Wireless Scattering Communication Networks}

\author{Guanchu Wang, Chen Gong, Zhimeng Jiang and Zhengyuan Xu
\thanks{This work was supported by Key Program of National Natural Science
Foundation of China (Grant No. 61631018) and Key Research Program of Frontier
Sciences of CAS (Grant No. QYZDY-SSW-JSC003).
}
\thanks{The authors are with Key Laboratory of Wireless-Optical Communications, Chinese Academy of Sciences, School of Information Science and Technology,
University of Science and Technology of China, Hefei, China.
Email: \{hegsns, zhimengj\}@mail.ustc.edu.cn, \{cgong821, xuzy\}@ustc.edu.cn.}
}

\newtheorem{Lemma}{Lemma}
\newtheorem{Theorem}{Theorem}

\begin{document}

\begin{spacing}{1.4}

\maketitle

\begin{abstract}
We propose a double-station access protocol (DS-CSMA) with multiple backoff mechanism for optical wireless scattering communication networks (OWSCN).
%, where two stations can transmit data to single destination simultaneously.can avoid the frames colliding with each other.
Furthermore, we extend existing Bianchi Markov model into state transmission model to analyze the collision probability, throughput and average delay.
For the application of protocol, we propose to optimize the initial contention window and indicator matrix to maximize throughput.
Both numerical and simulation results imply that the proposed protocol can achieve higher throughput and lower transmission delay compared with state-of-art baseline.
\end{abstract}

\begin{IEEEkeywords}
OWSCN, DS-CSMA, multiple backoff mechanism, collision probability, throughput, average delay.
\end{IEEEkeywords}

\section{Introduction}

Optical wireless scattering communication (OWSC) can potentially offer high data rate transmission due to its large bandwidth \cite{gagliardi1976optical, sevincer2013lightnets}.
Without emitting or being negatively affected by electromagnetic radiation, it can be applied to many scenarios where conventional radio-frequency (RF) communication is prohibited, for instance in the battlefield where radio silence is required \cite{xu2008ultraviolet}.
Various physical layer techniques has been proposed as the foundation of optical wireless scattering communication networks (OWSCN), including the multi-user signal detection \cite{wang2018signal, xiao2018Channel}, information security \cite{zou2018secrecy}, neighbor discovery \cite{li2011neighbor} and error correction codes \cite{wang2018study}.

\textcolor{red}{
Different from RF-based communication networks, the received signal of OWSC exhibits the characteristics of discrete photoelectrons due to the extremely large path loss \cite{Ding2009}.
Hence, Poisson-type channel model has been adopted for photon-counting-based physical-layer signal processing \cite{frey1991information, lapidoth2008capacity}.
Network communication for OWSC has been studied in existing works \cite{shaw2007hybrid}.
Specifically, a cluster-based algorithm has been proposed in \cite{li2011neighbor} for neighbour discovery;
the achievable rates of multiple access users have been addressed in \cite{wang2018signal};
%the performance of multi-packet reception (MPR) has been analyzed in \cite{Babich2010};
and a count-and-forward protocol has been proposed in \cite{gong2014non}.
%Since visible light communication suffers from misalignment, which may badly limit the throughput of optical wireless networks, we consider to employ OWSC as the physical-layer technology in network communication.
Furthermore, according to \cite{wang2018characterization, Guanchu8845642}, the authors proposed a superimposed transmission for OWSC, where different users are assigned into different signal layers so that the overall symbol rate doubles without reducing the symbol duration, and investigated the physical-layer techniques of transmission such as channel parameter estimation, signal detection and decoding.
}
%Techniques for OWSCN has been studied for years in existing works, where ultraviolet (UV) light communication is employed as the physical-layer technology.
%, mesh networks structure for FSO has been considered to analyze the propagation, directionality and mobility of FSO;

% \subsection{Related Literature Review}

\textcolor{red}{
As a classical medium access control (MAC) protocol, Carrier Sense Multiple Access (CSMA) is a widely adopted in wireless networks and leads to many improved versions \cite{lu2012survey}.
One typical method is to combine Carrier Sense Multiple Access with Collision Avoidance (CSMA/CA) with multi-packet reception (MPR) \cite{gau2009modeling}.
Specifically, MPR protocol employs CSMA/CA protocol and Request To Send/Clear To Send (RTS/CTS) mechanism for medium access control, and introduces Code Devision Multiple Access(CDMA) or Orthogonal Frequency Division Multiple Access (OFDMA) to transmit multiple data packages \cite{Babich2010}.
It allows higher throughput transmission than the conventional CSMA and gradually leads to many advanced versions which can be applied in different scenarios,
such as MPR with random access \cite{Dua2008}, acknowledgment-aware asynchronous patterns \cite{mukhopadhyay2013design} and adaptive Backoff Algorithm \cite{venkatesh2016qos}.
}

\textcolor{red}{
Even through MPR enables higher throughput by employing CDMA or OFDMA into the protocol, the overall throughput is still limited by the single backoff mechanism of conventional CSMA/CA.
In this work, we discover that multiple backoff mechanism results in higher throughput than the conventional single backoff mechanism in CSMA, and propose double-station CSMA (DS-CSMA) protocol for OWSCN.
% Similar to MPR, we assume double stations can simultaneously transmit superimposed data packages without interference with each other.
Different from MPR, without depending on CDMA, OFDMA or other orthogonal multi-user communication, we adopts superimposed transmission \cite{Guanchu8845642}
to deal with the packages from double stations in the physical-layer, where  channel estimation, synchronization, joint detection and decoding have been investigated in \cite{wang2018characterization}.}

\textcolor{red}{
Such MAC protocol with multiple backoff mechanism and superimposed transmission has not been addressed for radio-frequency wireless communication yet.
% and can achieve higher throughput than state-of-art.
%Our proposed DS-CSMA protocol can maximize the network throughput than existing works.
Specifically, as shown in Figure \ref{fig:MTC_illus}, each station employs multiple time counter to execute backoff processes.
%for the chance of data transmission.
Any two time counters (marked with the same color) are regarded as a pair, and one of the time counter backoff to 0 can trigger the pairs' superimposed transmission.
Such multiple backoff mechanism is more flexible than conventional single backoff mechanism in maximizing overall throughput and reducing transmission delay.
}
%, since optimal groups of time counters can be found to maximize , which will be one of our contributions in this work.  for optimization

%in the pattern double station accessing
%The pairs of stations than can transmit data simultaneously are characterized in an indicator matrix which depends on the physical-layer achievable rate.
%each station employ multiple backoff processes to
%
%Specifically, as shown in Figure \ref{fig:adhoc_config}, two stations can transmit their DATA frames to a common destination in asynchronous patterns, and the destination can successfully receive the DATA frames by joint detection and decoding.
%Meanwhile, the data collision can be avoidable as much as possible via the exchanges of assisted frames.
%The pairs of stations than can transmit data simultaneously are characterized in an indicator matrix which depends on the physical-layer achievable rate.

\begin{figure}
\subfigure[Time counter $(2, 1)$ backoff to zeros.]{
  \begin{minipage}[t]{0.34\linewidth}
    \centering
    \includegraphics[width=0.95\textwidth]{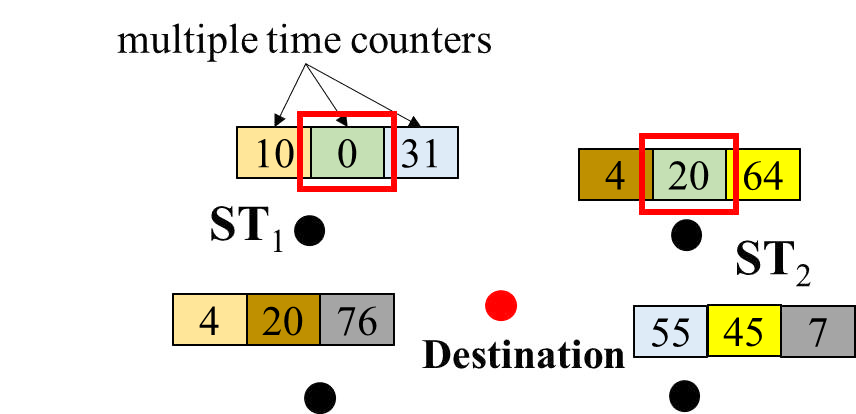}
    \label{fig:MTC1}
  \end{minipage}
}
\subfigure[Superimposed transmission from stations 1 and 2.]{
  \begin{minipage}[t]{0.26\linewidth}
    \centering
    \includegraphics[width=0.95\textwidth]{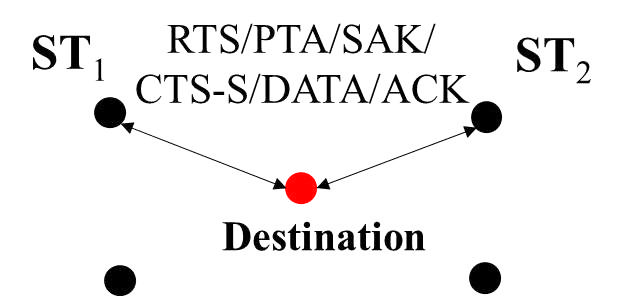}
    \label{fig:MTC2}
  \end{minipage}
}
\subfigure[Time counters $(2,1)$ and $(1,2)$ randomly take new values for next backoff.]{
  \begin{minipage}[t]{0.32\linewidth}
    \centering
    \includegraphics[width=0.95\textwidth]{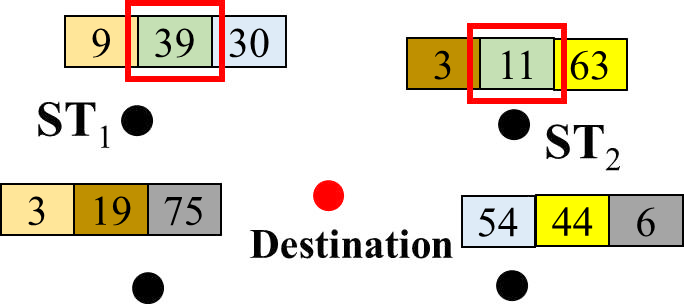}
    \label{fig:MTC3}
  \end{minipage}
}
\caption{\label{fig:MTC_illus} \textbf{Multiple backoff mechanism:} Each station employs multiple backoff process to strive for the chance of data transmission. Each lattice represents a time counter to execute the backoff process; 2 lattices with the same color implies the 2 stations will have superimposed transmission when backoff to 0; Time counter $(i,j)$ means time counter $i$ of station $j$.}

\end{figure}

\textcolor{red}{
On the other hand, Bianchi Model (based on discrete Markov model) has been widely applied in throughput analysis for various types of MAC protocol, e.g. \cite{Babich2010, mukhopadhyay2013design, bianchi2000performance}.
%, and leads to many progressive versions to analyze other
However, existing works on Bianchi Model cannot be directly applied to our proposed DS-CSMA protocol, since the state transition depends on more than one stations.
Hence, we extend Bianchi Model to the state transition model considered in this work, and obtain the numerical solutions to the throughput and collision probability.
Such numerical results are validated by simulation results.
Furthermore, we propose to approximately maximize the throughput with respect to the initial contention window and indicator matrix.
Numerical results demonstrate that the proposed protocol with optimized parameters can significantly outperform CSMA-based MAC including that with MPR.
}

The remainder of this paper is organized as follows.
In Section II, we introduce the indicator matrix and specify the DS-CSMA protocol.
In Sections III and IV, we provide a state transition model to analyze the collision probability and throughput.
In Section V, we propose to optimize the initial contention window and indicator matrix to maximize throughput.
Numerical and simulation results are shown in Section VI to evaluate the performance of DS-CSMA protocol, and the comparison with conventional CSMA including that with MPR.
% compared with \textcolor{red}{state-of-art baseline}.
Finally, Section VII concludes this work.

%\section{Related Works}
%
%Multi-Packet Reception

%\textcolor{red}{
%	However, existing works have failed to consider the uniqueness of OWSC in designing the protocol of medium access control (MAC) \cite{1186545, 7060515}.
%	Due to the extremely large path loss of OWSC, its received signal exhibits the characteristics of discrete photons, which cannot employ FDMA-based multi-channel CSMA for MAC.
%	Furthermore, MPR \cite{lu2012survey} depends on physical layer technologies like orthogonal CDMA, MIMO or space-time codes, which either reduce the transmission rate, or boosts the detection complexity of OWSC.
%}

\section{Double Station CSMA Protocol with Multiple Backoff Mechanism}

\subsection{Superimposed Frame Detection with Symbol Boundary Misalignment for OWSCN}

We consider an OWSCN consisting of $N_s$ active stations, where multiple stations can transmit packages to a common destination (only single destination is considered in this work).
Since different stations' transmissions can be performed in asynchronous patterns and interfere with each other, the frames can be superimposed with symbol boundary misalignment, where the relative delays are within one symbol duration.
\textcolor{red}{The achievable rate and frame detection for the superimposed transmission have been investigated in \cite{wang2018characterization, Guanchu8845642}, which can be adopted as the physical-layer technique to transmit data frames from two stations considered in this work.}
In this work, we propose a double-station access protocol based on CSMA/CA that can utilize the joint detection of superposed signal from two different stations.

In the protocol, each station $i$ deploys $J_i$ independent time counters $T_{i,1}, T_{i,2}, \cdots, T_{i,J_i}$ to control the packet transmission.
For the convenience of specification, we define the following two important terms and other notations in Table \ref{tb:symbol_definition_station}.
\begin{itemize}
\item[$\bullet$] \textbf{Time counter pair} (TCPair): The two time counters in two stations that can  transmit simultaneously with signal superposition at the receiver.
\item[$\bullet$] \textbf{Partner time counter} (PTCounter): Within a TCPair, the two time counters are PTCounter of each other.
\end{itemize}

%In this section, we characterize the key parameters and specify the proposed protocol.

%$\delta_1, \delta_2, ..., \delta_{J_i}$ can be obtained in PCF.

\subsection{Indicator matrix}

An indicator matrix $\boldsymbol{\Phi}$ is assumed known for the destination to reserve all indices of TCPairs, where $\boldsymbol{\Phi} = [\phi_{i,j} | 1 \leq i \leq N_s, 1 \leq j \leq N_s]$ is a zero-one matrix depending on the link gains in the physical layer.
We let $J_i = \sum^{N_s}_{k=1} \phi_{i,k}$ denote the number of PTCounters with station $i$, and $N$ denote the total number of TCPairs in the networks, where $N = \frac{1}{2}\sum_{i=1}^{N_s} J_i = \frac{1}{2}\sum_{i=1}^{N_s}\sum_{i'=0}^{N_s} \phi_{i,i'}$.
We use $\phi_{i,i'} = 1$ to indicate that stations $i$ and $i'$ can have a superimposed transmission of data frames controlled by TCPair $(T_{i,j}, T_{i',j'})$, where $j = \sum^{i'}_{k=1} \phi_{i,k}$ and $j' = \sum^i_{k=1} \phi_{i',k}$;
$\boldsymbol{\Phi}$ is symmetrical and $\phi_{i,i} = 0$ since TCPairs $(T_{i,j},T_{i',j'})$ and $(T_{i',j'}, T_{i,j})$ must coexist for $i \neq i'$.
An example is that
\begin{equation}
\begin{aligned}
\label{eq:partner_map1}
\boldsymbol{\Phi} = \left[
\begin{smallmatrix}
0&1&0&1&1 \\
1&0&1&1&1 \\
0&1&0&1&1 \\
1&1&1&0&0 \\
1&1&1&0&0
\end{smallmatrix}
\right],
%\quad \quad
%\widetilde{\boldsymbol{\Phi}} = \left\{
%\begin{smallmatrix}
%\{ 2,&4,&5 \} \\
%\{ 1,&3,&4,&5 \} \\
%\{ 2,&4,&5 \} \\
%\{ 1,&2,&3 \} \\
%\{ 1,&2,&3 \}
%\end{smallmatrix}
%\right\},
\end{aligned}
\end{equation}
where $\phi_{1,2} = \phi_{2,1} = 1$ indicates $T_{1,1}$ of station 1 and $T_{2,1}$ of station 2 can form TCPair $(T_{1,1}, T_{2,1})$; and $T_{1,1}$ and $T_{2,1}$ are PTCounters of each other.

\subsection{Channel Contention with Multiple Backoff Mechanism}

Before transmission, the initial values of all time counters $T_{i,j}$ for $1 \leq i \leq N_s, 1 \leq j \leq J_i$ are uniformly chosen from $[0, W_{i,j}-1]$, where $W_{i,j}$ denotes the contention window of $T_{i,j}$.
A common minimum and maximum contention window $W_0$ and $W_{\max}$ are shared by all time counters.
% maximum contention window

\begin{table}[tbp]
\caption{Specification of notations.}
\label{tb:symbol_definition_station}
\centering
\begin{tabular}{|c|c|c|}  % {lccc}
\hline
& Notation & Denotation\\
\hline
\multirow{3}{*}{OWSCN} & $N_s$ & Number of stations \\
\cline{2-3}
& $\boldsymbol{\Phi}$ & \textcolor{red}{Indicator matrix} \\
\cline{2-3}
& $N$ & Number of TCPairs \\
\hline
\multirow{5}{*}{Station $i$} & $J_i$ & Number of time counters \\
\cline{2-3}
& $T_{i,1}, T_{i,2}, \cdots, T_{i,J_i}$ & Time counters \\
\cline{2-3}
& $W_{i,1}, W_{i,2}, \cdots, W_{i,J_i} $ & Current contention window \\
\cline{2-3}
& $W_0$ & Minimum contention window \\
\cline{2-3}
& $W_{\max}$ & Maximum contention window \\
\hline
\end{tabular}
\end{table}

For each time counter $T_{i,j}$, $W_{i,j}$ is selected as $2^{m_{i,j}} W_0$, where $m_{i,j}$ takes from $0,1,2,\cdots,M-1$ with $W_{\max} = 2^{M-1} W_0$.
Furthermore, the time counters are decremented as long as the channel is sensed idle, as the \textbf{backoff} process in the CSMA/CA protrocol, and keep unchanged when the channel is sensed busy.
More specifically, the proposed protocol is characterized into the following $3$ parts.

\begin{itemize}
\item[1)]Once a time counter $T_{i,j}$ of station $i$ decreases to $0$, its backoff process is stopped, and station $i$ conducts the following active transmission:
\begin{itemize}
\item[1a)] It sends the RTS frame (request to send frame) to the destination, which contains its 2-dimensional index $(i,j)$, as shown in Figure \ref{fig:flowchart}.4., and waits for the PTA frame (partner activate) from the destination.

\item[1b)] Once receiving the PTA frame, it waits for the CTS-S frame (clear to send-superimposed transmission) from the destination.
    In case of missing the PTA frame (waiting for the period of PTA frame but failing to receive the PTA frame), it doubles the contention window of $T_{i,j}$ by updating $W_{i,j} \gets 2 W_{i,j}$, as shown in Figure \ref{fig:flowchart}.8., and jump to step \emph{1e)}.

\item[1c)] After receiving the CTS-S, it transmits data frame to the destination, as shown in Figure \ref{fig:flowchart}.14., and waits for ACK (acknowledgement) from the destination.

\item[1d)] On receiving the ACK, it resets the contention window of $T_{i,j}$ to $W_0$, i.e., $W_{i,j} \gets W_0$.

\item[1e)] Randomly choose an integer from $[0,W_{i,j}-1]$ for $T_{i,j}$ as its initial backoff value, and restarts the backoff process in a DIFS.
    Other time counters also reactive their previous backoff process in a DIFS.
\end{itemize}
\end{itemize}

\begin{itemize}
\item[2)]Once the destination receives the RTS frame from a certain station with the 2-dimensional index $(i,j)$ of $T_{i,j}$, it obtains the index $(i',j')$ of PTCounter $T_{i',j'}$ according to the indicator matrix $\boldsymbol{\Phi}$, and manages to receive data frames from stations $i$ and $i'$ into the following steps:
\begin{itemize}
\item[2a)] It broadcasts the PTA frame containing $(i',j')$ to all stations, and waits for the SAK (superposition acknowledgement) from station $i'$. % as shown in Figure \ref{fig:frame_stream},

\item[2b)] Once receiving the SAK, it broadcasts the CTS-S frame to stations $i$ and $i'$, as shown in Figure \ref{fig:frame_stream}, and waits for the superimposed data transmission.

\item[2c)] After successfully decoding the superimposed data frames, it broadcasts the ACK to all stations. %, as shown in Figure \ref{fig:frame_stream}.
\end{itemize}
\end{itemize}

\begin{itemize}
\item[3)] For station $\tilde{i} \neq i'$, upon receiving the PTA frame from the destination, its time counters stop backoff process such that the channel is available for stations $i$ and $i'$.
Once receiving the ACK from the destination, its time counters reactive their previous backoff process in a DIFS.
For station $i'$, upon receiving the PTA frame from the destination, its time counters stop the backoff process, and conducts the transmission via the following steps:

\begin{itemize}
\item[3a)] As shown in Figure \ref{fig:flowchart}.12., it transmits the SAK frame to the destination, which contains whether it will have a superimposed transmission with station $i$.
In case of refusing to have superimposed transmission with station $i$, it jump to \emph{3d)}.

\item[3b)] After transmitting the SAK frame, it waits for the CTS-S frame from the destination.

\item[3c)] Upon receiving the CTS-S frame, without having to cooperate with station $i$, it transmits data frame as soon as possible, as shown in Figure \ref{fig:flowchart}.14., and waits for the ACK from the destination.
%completes a superimposed transmission of data frame with station $i$ to the destination

\item[3d)] Once receiving the ACK, $T_{i',j'}$ resets the contention window to $W_0$, randomly chooses a value from $[0,W_0-1]$ as its initial backoff value, and restarts the backoff process in a DIFS.
    Other time counters also reactive their previous backoff process in a DIFS.

\end{itemize}
\end{itemize}

For clarification, we summarize all types of control frames in Table \ref{tb:frame_definition}.
A successful superimposed transmission of data frames consists of the exchanges of RTS, PTA, SAK, CTS, Data and ACK frames, as shown in Figure \ref{fig:frame_stream}.
A collision happens when more than one stations transmits RTSs to the destination simultaneously.
%, as shown in Figure \ref{fig:frame_stream_collision}.

\begin{table}[tbp]
\caption{Specification of different type of frames.}
\label{tb:frame_definition}
\centering
\begin{tabular}{|c|c|}  % {lccc} ��ʾ����Ԫ�ض��뷽ʽ��left-l,right-r,center-c
\hline
Frame type %& Full name
& Function \\
\hline
RTS %& Request to send
& Ask for occupying the common channel \\
\hline
PTA %& Partner activate
& Activate the PTCounter and freeze other TCPairs \\
\hline
SAK %& Superimposed transmission acknowledge
& Acknowledgement of a superimposed transmission \\
\hline
CTS-S %& Data frame trigger
& Clear to send-superimposed transmission \\
\hline
ACK %& Acknowledgement
& Flag of a complete superimposed transmission \\
\hline
\end{tabular}
\end{table}

%\begin{figure}
%\subfigure[Transmission of RTS.]{
%  \begin{minipage}[t]{0.3\linewidth}
%    \centering
%	\includegraphics[width=1.0\textwidth]{RTS.pdf}
%    \label{fig:RTS}
%  \end{minipage}%
%}
%\subfigure[Transmission of PTA.]{
%  \begin{minipage}[t]{0.3\linewidth}
%    \centering
%    \includegraphics[width=1.0\textwidth]{PTA.pdf}
%    \label{fig:PTA}
%  \end{minipage}
%}
%\subfigure[Transmission of SAK.]{
%  \begin{minipage}[t]{0.3\linewidth}
%    \centering
%    \includegraphics[width=1.0\textwidth]{SAK.pdf}
%    \label{fig:SAK}
%  \end{minipage}
%}
%$\\$
%\subfigure[Transmission of CTS.]{
%  \begin{minipage}[t]{0.3\linewidth}
%    \centering
%    \includegraphics[width=1.0\textwidth]{CTS.pdf}
%    \label{fig:CTS}
%  \end{minipage}
%}
%\subfigure[Superimposed transmission of data frames.]{
%  \begin{minipage}[t]{0.3\linewidth}
%    \centering
%    \includegraphics[width=1.0\textwidth]{DATA.pdf}
%    \label{fig:DATA}
%  \end{minipage}
%}
%\subfigure[Transmission of ACK.]{
%  \begin{minipage}[t]{0.3\linewidth}
%    \centering
%    \includegraphics[width=1.0\textwidth]{ACK.pdf}
%    \label{fig:ACK}
%  \end{minipage}
%}
%\caption{\label{fig:access_protocol} Illustration of a successful superimposed transmission.}
%\end{figure}

\begin{figure}
\centering
	\includegraphics[width=0.6\textwidth]{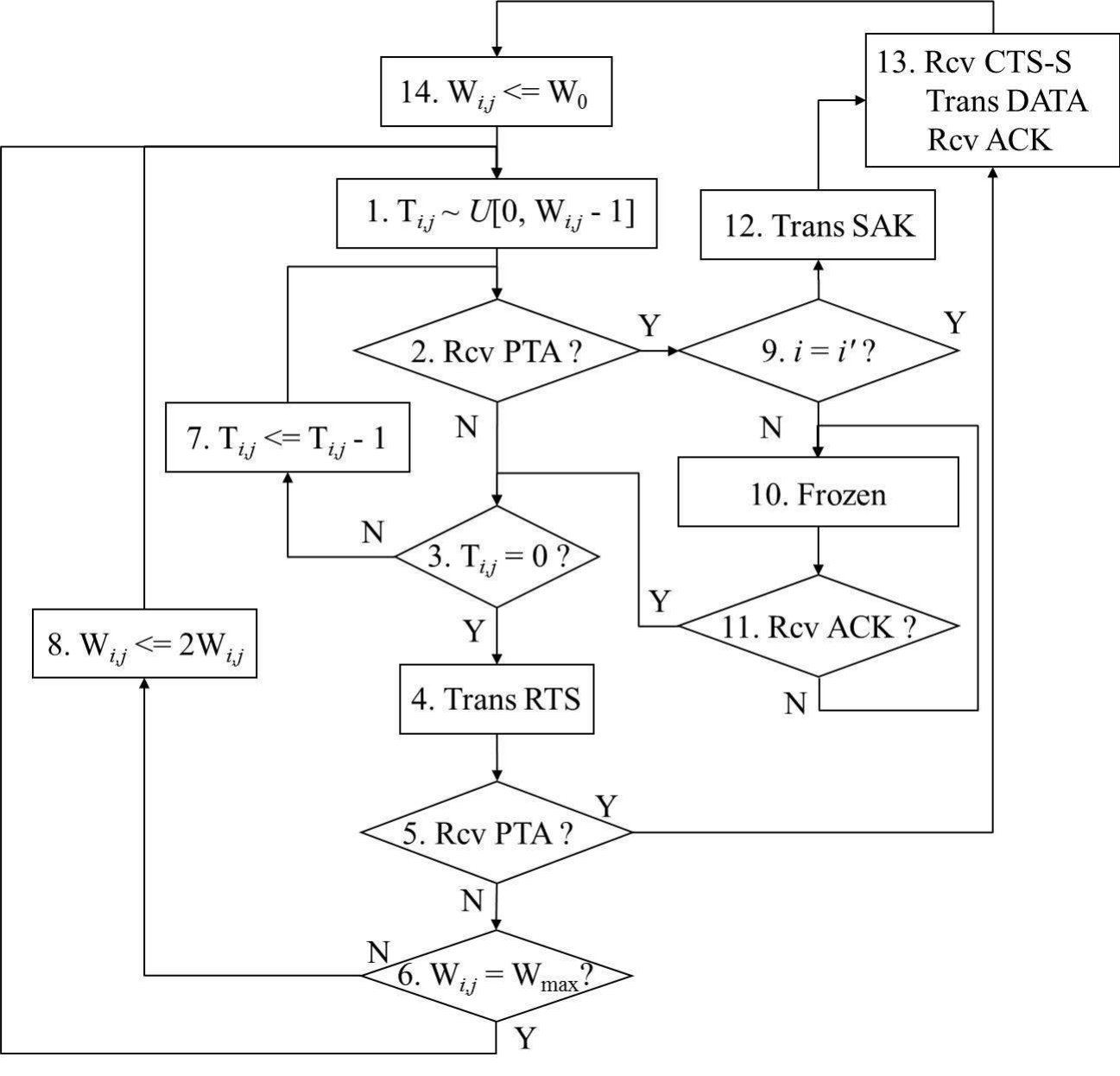}
    \caption{\label{fig:flowchart} Flowchart of DS-CSMA protocol.}
\end{figure}

\begin{figure}
\centering
	\includegraphics[width=0.9\textwidth]{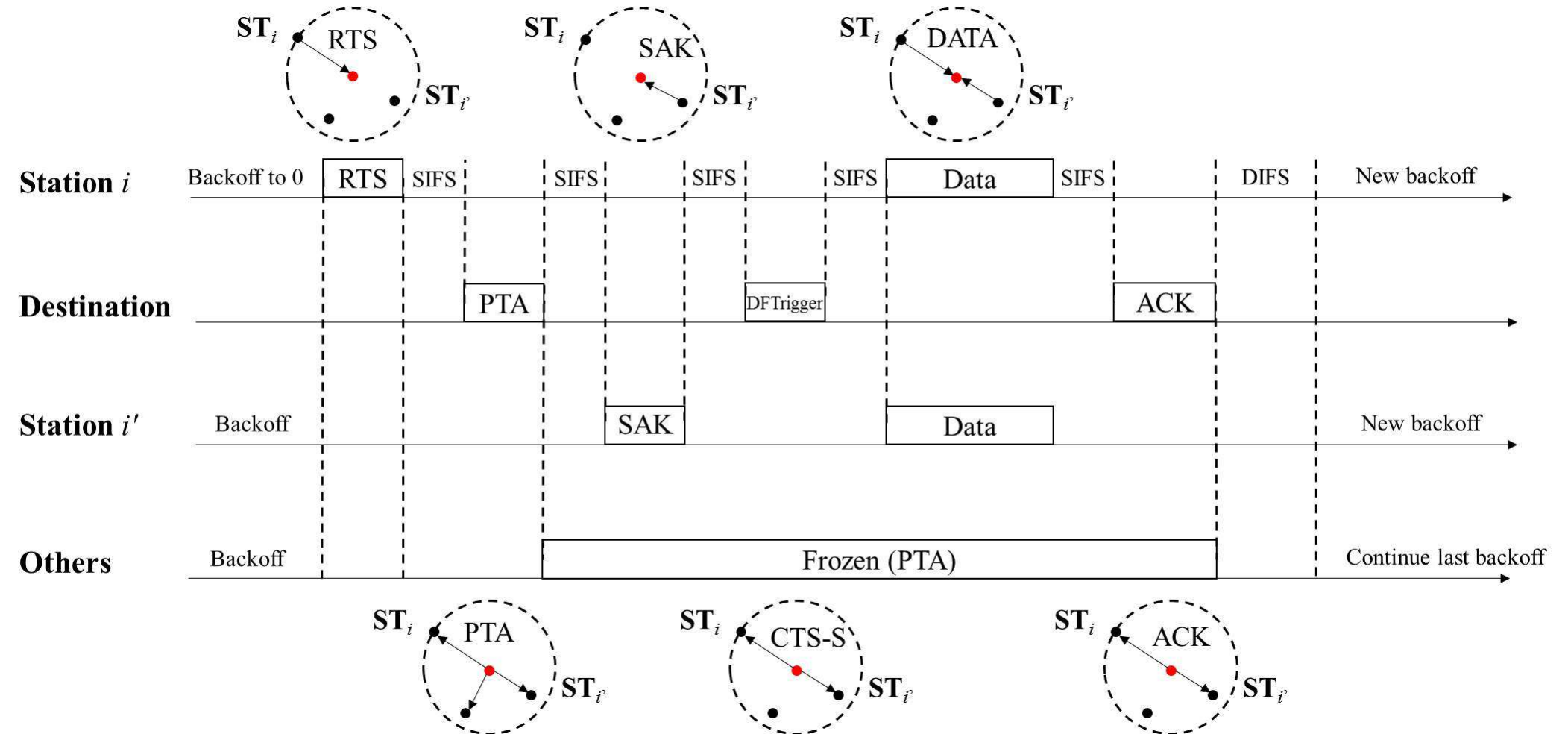}
    \caption{\label{fig:frame_stream} Frame sequence of a successful superimposed transmission.}
\end{figure}

%\begin{figure}
%\centering
%	\includegraphics[width=0.65\textwidth]{frame_stream_collision.pdf}
%    \caption{\label{fig:frame_stream_collision} Frame sequence of a collision.}
%\end{figure}

%\begin{figure}
%\centering
%	\includegraphics[width=0.69\textwidth]{BCSMA_algorithm.pdf}
%    \caption{\label{fig:BCSMA_algorithm} Diagram of DS-CSMA.}
%\end{figure}

\subsection{An Example of DS-CSMA}

We give an example on the proposed DS-CSMA protocol in this subsection.
We assume $5$ stations with the indicator matrix in Equation (\ref{eq:partner_map1}), where $8$ TCPairs are deployed to control data transmission.
Provided that $W_{i,j} = W_0 = 32$ and $W_{\max} = 128$, the  states of $8$ TCPairs at certain time point are shown in the first column of Table \ref{tb:example}.

After $3$ time slots, the TCPairs are illustrated in the second column of Table \ref{tb:example}, where $T_{1,2}$ and $T_{4,3}$ are both reduced to $0$.
Accordingly, stations $1$ and $4$ both transmit the RTS-S frame to the destination, where a collision happens.
% as shown in Figure \ref{fig:frame_stream_collision}.
Then, they conduct steps \emph{1a), 1b)} and \emph{1e)}.
The new initial backoff values are chosen from $[0,64]$, e.g., $T_{1,2} = 26, T_{4,3} = 58$.

After another $2$ time slots, the TCPairs are self-decrement twice and turn to the states shown in the third column of Table \ref{tb:example}, where only $T_{3,2}$ is reduced to $0$.
Therefore, stations $3$ and $4$ conduct a successful superimposed transmission as shown in Figure \ref{fig:frame_stream}, where station $3$ conducts steps \emph{1a)-1e)}; and station $4$ conducts steps \emph{3a)-3d)}.

\begin{table}[tbp]
\caption{Example of DS-CSMA.}
\label{tb:example}
\centering
\begin{tabular}{|c|c|c|}  % {lccc} ��ʾ����Ԫ�ض��뷽ʽ��left-l,right-r,center-c
\hline
Initial & $3$ time slots later & $5$ time slots later \\
\hline
\tabincell{c}{$(T_{1,1},T_{2,1})=(3,28)$\\ $(T_{1,2},T_{4,1})=(19,13)$\\ $(T_{1,3},T_{5,1})=(8,30)$\\ $(T_{2,2},T_{3,1})=(24,6)$\\ $(T_{2,3},T_{4,2})=(9,31)$\\ $(T_{2,4},T_{5,2})=(11,17)$\\ $(T_{3,2},T_{4,3})=(5,33)$\\ $(T_{3,3},T_{5,3})=(29,7)$} & \tabincell{c}{$(T_{1,1},T_{2,1})=(0,25)$\\ $(T_{1,2},T_{4,1})=(16,10)$\\ $(T_{1,3},T_{5,1})=(5,27)$\\ $(T_{2,2},T_{3,1})=(21,3)$\\ $(T_{2,3},T_{4,2})=(6,28)$\\ $(T_{2,4},T_{5,2})=(8,14)$\\ $(T_{3,2},T_{4,3})=(2,0)$\\ $(T_{3,3},T_{5,3})=(26,4)$} & \tabincell{c}{
$(T_{1,1},T_{2,1})=(24,23)$\\ $(T_{1,2},T_{4,1})=(14,8)$\\ $(T_{1,3},T_{5,1})=(3,25)$\\ $(T_{2,2},T_{3,1})=(19,1)$\\ $(T_{2,3},T_{4,2})=(4,26)$\\ $(T_{2,4},T_{5,2})=(6,12)$\\ $(T_{3,2},T_{4,3})=(0,56)$\\ $(T_{3,3},T_{5,3})=(24,2)$} \\
\hline
\end{tabular}
\end{table}

\section{State Transition Model for DS-CSMA}

\textcolor{red}{
Conventional theories (Bianchi Model) \cite{bianchi2000performance} of throughput analysis for distributed CSMA/CA cannot be directly applied to our proposed DS-CSMA protocol since each station depends on more than one time counters.
To address this issue, we choose to use state $(T_{i_1,j_1}, T_{i_2,j_2})$ to characterize a TCPair.
% Since different TCPairs' backoff process are independent with each other,
For simplicity but without loss of generality, we consider the state transition of certain TCPair as the representation of whole system.
%we can choose to investigate certain TCPair instead of the whole system.
The subscript of time counters $(T_{i_1,j_1}, T_{i_2,j_2})$ can be simplified into $(T_1, T_2)$ characterized by parameters $(m,n,i,j)$, where $m$ and $n$ denote that the contention windows of the  $T_1$ and $T_2$ equal $W_m =  2^{m} W_0$ and $W_n =  2^{n} W_0$, respectively; and $i$ and $j$ notate that $(T_1, T_2) = (i,j)$.}
Furthermore, let $\mathbb{P} (m,n,i,j)$ denote the probability of state $(m,n,i,j)$, and $\mathbb{P} (m,n,i,j | m',n',i',j')$ denote the state transition probability from state $(m',n',i',j')$ to state $(m,n,i,j)$, where $0 \leq m,m',n,n' \leq M-1, 0 \leq i,i' \leq W_m-1, 0 \leq j,j' \leq W_n-1$.

% We characterize the state and state transition probabilities in the remainder of this Section.

\subsection{State Transition Probabilities}

First of all, for $0 \leq i \leq W_m-1, 0 \leq j \leq W_n-1$, both $T_1$ and $T_2$ must be reduced by one in the next slot.
Hence the transition probability is given by
\begin{equation}
\begin{aligned}
\label{eq:transition_probability1}
\mathbb{P}(m,n,i,j | m,n,i+1,j+1) = 1.
\end{aligned}
\end{equation}

The remaining cases depend on parameter $p$, the probability that a collision happens in the channel due to other TCPairs' contention.
Probability $p$ is determined by parameters $N$ and $\eta$ as follows,
\begin{equation}
\begin{aligned}
\label{eq:collision_pro_eq}
p = 1 - (1 - \eta)^{N-1},
\end{aligned}
\end{equation}
where $N = \frac{1}{2}\sum_{i=1}^{N_s}\sum_{j=1}^{N_s}\phi_{i,j}$ denotes the number of TCPairs as shown in Table \ref{tb:symbol_definition_station}; and
\begin{equation}
\begin{aligned}
\label{eq:eta}
\eta = \sum_{m=0}^{M-1} \sum_{n=0}^{M-1} \Big[ \sum_{i=1}^{W_m-1} \mathbb{P}(m,n,i,0) + \sum_{j=1}^{W_n-1} \mathbb{P}(m,n,0,j) \Big]
\end{aligned}
\end{equation}
denotes the overall probability that one of the time counters is reduced to $0$.

Secondly, for $i=0$ or $j=0$, it will lead to two cases of state-transmission.
A new successful transmission, consisting of RTS, PTA, SAK, CTS, Data frame superimposed transmission and ACK, may happen with probability $(1-p)$.
After a successful transmission, $T_1$ and $T_2$ are independently initialized uniformly in the range $[0, W_0 - 1]$.
Accordingly, the transition probabilities are given by
\begin{equation}
\begin{aligned}
\label{eq:transition_probability2}
\mathbb{P}(0,0,i,j | m,n,i',0) = \frac{1-p}{W_0^2},
\quad
\mathbb{P}(0,0,i,j | m,n,0,j') = \frac{1-p}{W_0^2}.
\end{aligned}
\end{equation}
Besides, the RTS-S may collide with that from other stations with probability $p$.
In this case, if $m,n \neq M-1$, either $T_1$ or $T_2$ may be uniformly chosen in range $[0, W_{m+1} - 1]$ or $[0, W_{n+1} - 1]$ with probability $p$, and thus the transition probabilities are given by
\begin{equation}
\begin{aligned}
\label{eq:transition_probability3}
\mathbb{P}(m+1,n,i,j | m,n,0,j+1) = \frac{p}{W_{m+1}},
\quad
\mathbb{P}(m,n+1,i,j | m,n,i+1,0) = \frac{p}{W_{n+1}}.
\end{aligned}
\end{equation}
Furthermore, if $m=M-1$ or $n=M-1$, the contention windows of either $T_1$ or $T_2$ will never become doubled in subsequent transmission, and the transition probabilities are given by
\begin{equation}
\begin{aligned}
\label{eq:transition_probability4}
\mathbb{P}(M-1,n,i,j | M-1,n,0,j+1) = \mathbb{P}(m,M-1,i,j | m,M-1,i+1,0) = \frac{p}{W_{M-1}}.
% &= \frac{p}{W_{M-1}}
\end{aligned}
\end{equation}
In addition, if $m = n = M-1$, neither $T_1$ or $T_2$ will be doubled the contention window in subsequent transmissions.
Hence, the transition probabilities are given by
\begin{equation}
\begin{aligned}
\label{eq:transition_probability5}
\mathbb{P}(M-1,M-1,i,j | M-1,M-1,0,j+1) = \mathbb{P}(M-1,M-1,i,j | M-1,M-1,i+1,0) = \frac{p}{W_{M-1}}.
%\\
%\mathbb{P}(M-1,M-1,i,j | M-1,M-1,i+1,0) &= \frac{p}{W_{M-1}}
\end{aligned}
\end{equation}

Finally, considering $i=j=0$, where $T_1$ and $T_2$ decrease to $0$ simultaneously, where a collision must happen.
For $m,n\neq M-1$, $T_1$ and $T_2$ will uniformly take values in $[0, W_{m+1} - 1]$ and $[0, W_{n+1} - 1]$, respectively.
The transition probability is given by
\begin{equation}
\begin{aligned}
\label{eq:transition_probability6}
\mathbb{P}(m+1,n+1,i,j | m,n,0,0) = \frac{1}{W_{m+1} W_{n+1}}.
\end{aligned}
\end{equation}
For $m = M-1$ or $n = M-1$, either $W_m$ or $W_n$ is never doubled since it is up to the maximum contention window $W_{\max}$.
Then, the transition probabilities are given by,
\begin{equation}
\begin{aligned}
\label{eq:transition_probability7}
\mathbb{P}(M-1,n+1,i,j | M-1,n,0,0) &= \frac{1}{W_{M-1} W_{n+1}},
\\
\mathbb{P}(m+1,M-1,i,j | m,M-1,0,0) &= \frac{1}{W_{m+1} W_{M-1}}.
\end{aligned}
\end{equation}
For $m = n = M-1$, neither $W_m$ or $W_n$ will be doubled, and the transition probabilities are given by
\begin{equation}
\begin{aligned}
\label{eq:transition_probability8}
\mathbb{P}(M-1,M-1,i,j | M-1,M-1,0,0) = \frac{1}{W_{M-1}^2}.
\end{aligned}
\end{equation}

\subsection{The State Probabilities}

We obtain the state probabilities in this subsection based on the state transition probabilities.
Since $i$ and $j$ take values in $[0, W_m-1]$ and $[0, W_n-1]$, respectively,
it is convinced that $\mathbb{P} (m,n,W_m,\bullet) = \mathbb{P} (m,n,\bullet,W_n) = \mathbb{P} (m,n,W_m,W_n) = 0$ for $0 \leq m,n \leq M-1$. We summarize all possible state probabilities into the following $7$ cases.

%For the convenience of discussion, but ,

\textbf{Case 1}: For $m=n=0, 0 \leq i \leq W_0-1, 0 \leq j \leq W_0-1$, as shown in Figure \ref{fig:m=0,n=0}, the previous states of $(0,0,i,j)$ are $(0,0,i+1,j+1)$, $(0,0,0,1), (0,0,0,2), \cdots, (M-1,M-1,0,W_{M-1}-1)$ as well as $(0,0,1,0), (0,0,2,0), \cdots, (M-1,M-1,W_{M-1}-1,0)$, and thus
%the state probability $\mathbb{P}(0,0,i,j)$ is given by
\begin{equation}
\begin{aligned}
\label{eq:ste_Pr1}
\mathbb{P}(0,0,i,j) = \mathbb{P}(0,0,i+1,j+1)+ \eta \frac{1-p}{W^2_0},
\end{aligned}
\end{equation}
where $\eta$ is given by Equation (\ref{eq:eta}).

\textbf{Case 2}: For $0 < m < M-1, n = 0, 0 \leq i \leq W_m-1, 0 \leq j \leq W_0-1$, as shown in Figure \ref{fig:n=0}, the previous states of $(m,0,i,j)$ are $(m,0,i+1,j+1)$ as well as $(m-1,0,0,j+1)$, and thus
%the state probability $\mathbb{P}(m,0,i,j)$ is given by
\begin{equation}
\begin{aligned}
\label{eq:ste_Pr2}
\mathbb{P}(m,0,i,j) = \mathbb{P}(m,0,i+1,j+1) + \frac{\mathbb{P}(m-1,0,0,j+1)}{W_m}p.
\end{aligned}
\end{equation}

\textbf{Case 3}: For $m = M-1, n = 0$, $0 \leq i \leq W_{M-1}-1$ and $1 \leq j \leq W_0-1$, as shown in Figure \ref{fig:m=M-1,n=0}, the previous states of $(M-1,0,i,j)$ are $(M-1,0,i+1,j+1)$, $(M-2,0,0,j+1)$ as well as $(M-1,0,0,j+1)$, and thus
%the state probability $\mathbb{P}(M-1,0,i,j)$ is given by
\begin{equation}
\begin{aligned}
\label{eq:ste_Pr3}
\mathbb{P}(M-1,0,i,j) = \mathbb{P}(M-1,0,i+1,j+1) + \frac{\mathbb{P}(M-2,0,0,j+1)}{W_{M-1}}p + \frac{\mathbb{P}(M-1,0,0,j+1)}{W_{M-1}}p.
\end{aligned}
\end{equation}

\textbf{Case 4}: For $0 < n \leq m < M-1$, $0 \leq i \leq W_m-1, 0 \leq j \leq W_n-1$, as shown in Figure \ref{fig:Markov_state} (d), the previous states of $(m,n,i,j)$ are $(m,n,i+1,j+1), (m-1,n-1,0,0), (m-1,n,i,j+1)$ as well as $(m,n-1,i+1,j)$, and thus
%the state probability $\mathbb{P}(m,n,i,j)$ is given by
\begin{equation}
\begin{aligned}
\label{eq:ste_Pr4}
\!\!\!\! \mathbb{P}(m,n,i,j) \!=\! \mathbb{P}(m,n,i+1,j+1)+ \frac{\mathbb{P}(m-1,n-1,0,0)}{W_m W_n} + \frac{\mathbb{P}(m-1,n,i,j+1)}{W_m} p + \frac{\mathbb{P}(m,n-1,i+1,j)}{W_n}p.
\end{aligned}
\end{equation}

\textbf{Case 5}: For $m = M-1, 0 < n < M-1$, $0 \leq i \leq W_{M-1}-1, 0 \leq j \leq W_n-1$, as shown in Figure \ref{fig:Markov_state} (e), the previous states of $(M-1,n,i,j)$ are $(M-2,n-1,0,0), (M-1,n-1,0,0), (M-2,n,0,j+1), (M-1,n-1,i+1,0)$ as well as $(M-1,n,0,j+1)$, and thus
%the state probability $\mathbb{P}(M-1,n,i,j)$ is given by
\begin{equation}
\begin{aligned}
\label{eq:ste_Pr5}
\mathbb{P}(M-1,n,i,j) &= \mathbb{P}(M-1,n,i+1,j+1) + \frac{\mathbb{P}(M-2,n-1,0,0)}{W_{M-1} W_n} + \frac{\mathbb{P}(M-1,n-1,0,0)}{W_{M-1} W_n}
\\
&+ \frac{\mathbb{P}(M-2,n,0,j+1)}{W_{M-1}}p + \frac{\mathbb{P}(M-1,n-1,i+1,0)}{W_n}p + \frac{\mathbb{P}(M-1,n,0,j+1)}{W_{M-1}}p
\end{aligned}
\end{equation}

\textbf{Case 6}: For $m = n = M-1$, $0 \leq i,j \leq W_{M-1}-1$, as shown in Figure \ref{fig:Markov_state} (f), the previous states of $(M-1,M-1,i,j)$ are $(M-1,M-1,i+1,j+1), (M-2,M-1,0,j+1), (M-1,M-2,i+1,0), (M-1,M-1,0,j+1), (M-1,M-1,i+1,0), (M-1,M-1,0,0), (M-2,M-2,0,0), (M-2,M-1,0,0),  (M-1,M-2,0,0)$ as well as $(M-1,M-1,0,0)$.
Therefore, we have
% and the state probability $\mathbb{P}(M-1,M-1,i,j)$ is given by
\begin{equation}
\begin{aligned}
\label{eq:ste_Pr6}
\mathbb{P}(M \!-\! 1,\! M \!-\! 1,\! i,\! j) &= \mathbb{P}(M \!-\! 1,\! M \!-\! 1,\! i+1,\! j \!+\! 1) + \frac{\mathbb{P}(M \!-\! 2,\! M \!-\! 1,\! 0,\! j \!+\! 1)}{W_{M-1}}p + \frac{\mathbb{P}(M \!-\! 1,\! M \!-\! 2,\! i \!+\! 1,\! 0)}{W_{M-1}}p
\\
&+ \frac{\mathbb{P}(M \!-\! 1,M \!-\! 1,0,j \!+\! 1)}{W_{M-1}}p + \frac{\mathbb{P}(M \!-\! 1,M \!-\! 1,i \!+\! 1,0)}{W_{M-1}}p + \frac{\mathbb{P}(M-1,M-1,0,0)}{W^2_{M-1}}
\\
&+ \frac{\mathbb{P}(M-2,M-2,0,0)}{W^2_{M-1}} + \frac{\mathbb{P}(M-2,M-1,0,0)}{W^2_{M-1}} + \frac{\mathbb{P}(M-2,M-1,0,0)}{W^2_{M-1}}
\end{aligned}
\end{equation}

\textbf{Case 7}: For $0 \leq m < n \leq M-1$, $0 \leq i \leq W_m-1$ and $0 \leq j \leq W_n-1$, we have
\begin{equation}
\begin{aligned}
\label{eq:ste_Pr7}
\mathbb{P}(m,n,i,j) = \mathbb{P}(n,m,j,i).
\end{aligned}
\end{equation}

%\emph{Proof:} Please refer to Appendix A.

\begin{figure}
\setlength{\abovecaptionskip}{0cm}
\setlength{\belowcaptionskip}{-10cm}
\subfigure[State transition of cases 1, $m=n=0$.]{
% $\qquad\qquad\qquad$
  \begin{minipage}[t]{0.55\linewidth}
    \centering
	\includegraphics[width=0.8\textwidth]{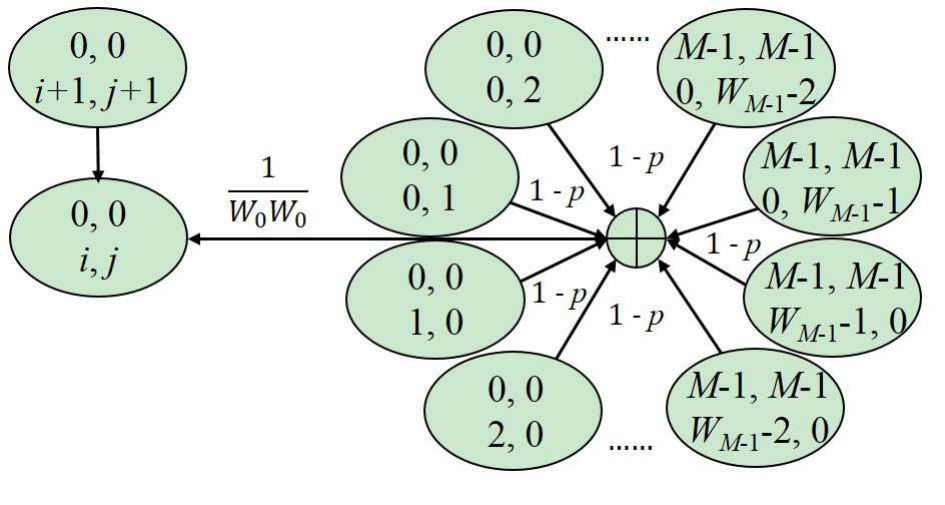}
    \label{fig:m=0,n=0}
  \end{minipage}%
}
\subfigure[State transition of case 2, $0<m<M-1,n=0$.]{
$\qquad\qquad\qquad$
  \begin{minipage}[t]{0.25\linewidth}
    \centering
    \includegraphics[width=0.8\textwidth]{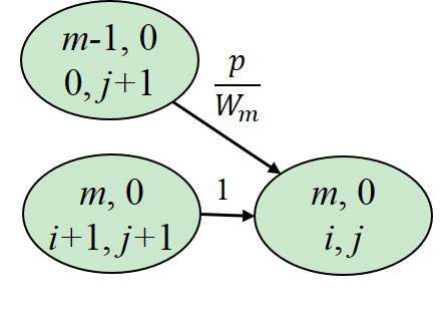}
    \label{fig:n=0}
  \end{minipage}
}
$\\$
\subfigure[State transition of case 3, $m=M-1,n=0$.]{
  \begin{minipage}[t]{0.3\linewidth}
    \centering
    \includegraphics[width=0.8\textwidth]{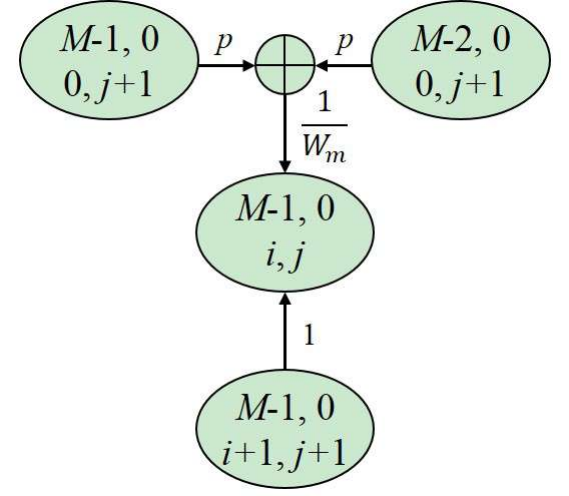}
    \label{fig:m=M-1,n=0}
  \end{minipage}
}
\subfigure[State transition of case 4, $0\leq m,n\leq M-1$.]{
  \begin{minipage}[t]{0.3\linewidth}
    \centering
    \includegraphics[width=0.8\textwidth]{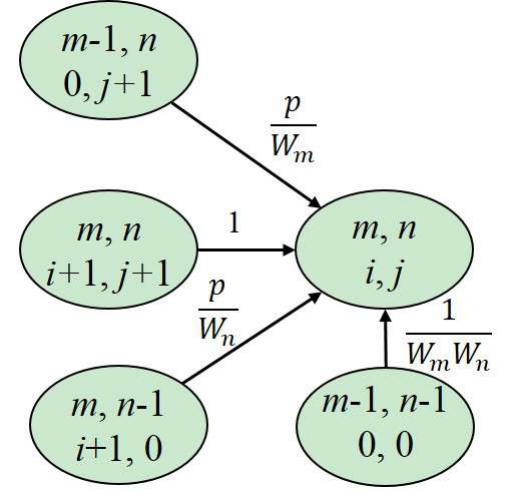}
    \label{fig:m,n}
  \end{minipage}
}
\subfigure[State transition of case 5, $0<m<M-1,n=M-1$.]{
  \begin{minipage}[t]{0.4\linewidth}
    \centering
    \includegraphics[width=0.8\textwidth]{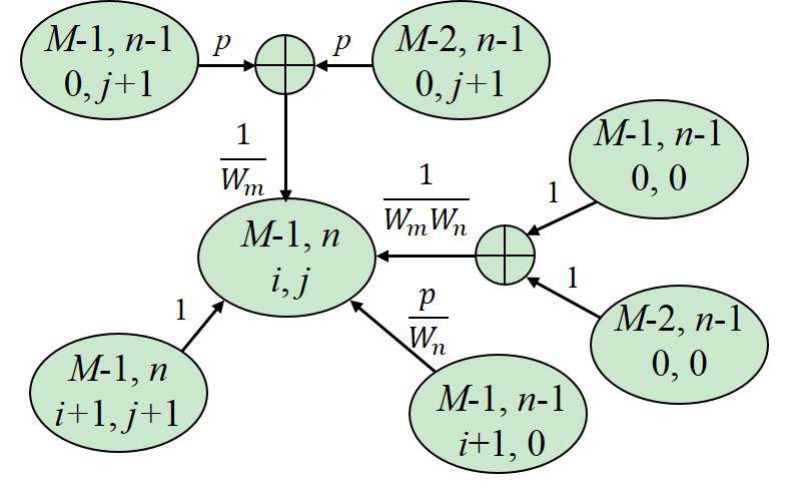}
    \label{fig:n=M-1}
  \end{minipage}
}
$\\$
\subfigure[State transition of case 6, $m=n=M-1$.]{
$\qquad\qquad\qquad\qquad\qquad$
  \begin{minipage}[t]{0.6\linewidth}
    \centering
    \includegraphics[width=0.8\textwidth]{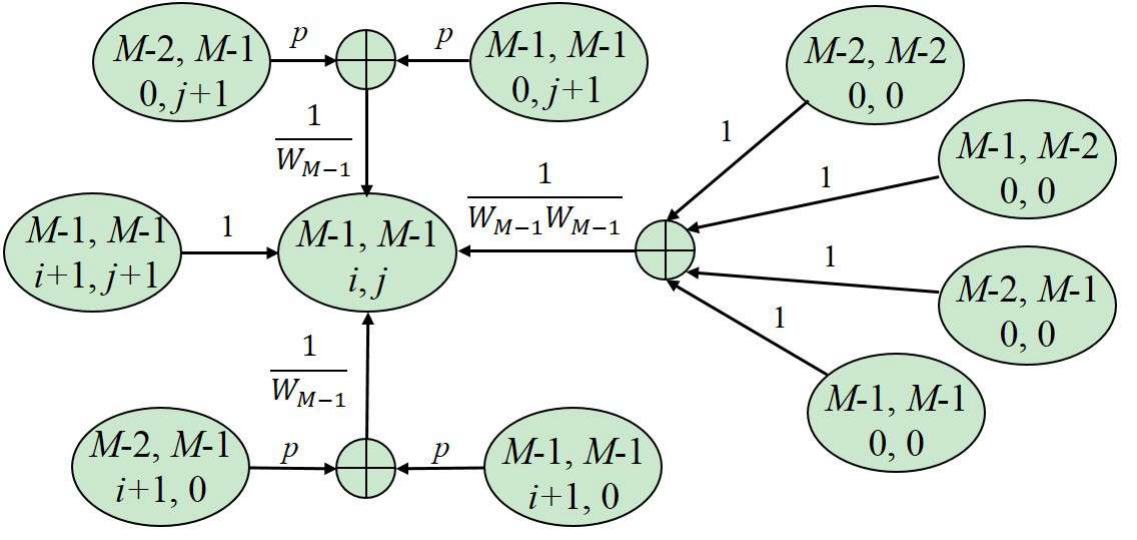}
    \label{fig:m=M-1,n=M-1}
  \end{minipage}
}
\caption{\label{fig:Markov_state}State transition of the cases 1-6.}
\end{figure}

%\begin{table}[tbp]
%\caption{Characteristics of the $7$ cases.}
%\label{tb:ste_pro}
%\centering
%\begin{tabular}{|c|c|c|c|c|}  % {lccc} ��ʾ����Ԫ�ض��뷽ʽ��left-l,right-r,center-c
%\hline
%Case & Specification & \tabincell{c}{Illustration of \\ state transition} & \tabincell{c}{Related state \\ transition probability} & State probability \\
%\hline
%1 & $m=n=0$ & Fig. \ref{fig:m=0,n=0} & (\ref{eq:transition_probability1}), (\ref{eq:transition_probability2}) & (\ref{eq:ste_Pr1}) \\
%\hline
%2 & $m=M-1,n=0$ & Fig. \ref{fig:n=0} & (\ref{eq:transition_probability1}) (\ref{eq:transition_probability3}) & (\ref{eq:ste_Pr2}) \\
%\hline
%3 & $0<m<M-1,n=0$ & Fig. \ref{fig:m=M-1,n=0} & (\ref{eq:transition_probability1}), (\ref{eq:transition_probability3}), (\ref{eq:transition_probability4}) & (\ref{eq:ste_Pr3}) \\
%\hline
%4 & $0<n\leq m<M-1$ & Fig. \ref{fig:m,n} & (\ref{eq:transition_probability1}), (\ref{eq:transition_probability3}), (\ref{eq:transition_probability6}) & (\ref{eq:ste_Pr4}) \\
%\hline
%5 & $m=M-1,0<n<M-1$ & Fig. \ref{fig:n=M-1} & (\ref{eq:transition_probability1}), (\ref{eq:transition_probability3}), (\ref{eq:transition_probability4}), (\ref{eq:transition_probability6}), (\ref{eq:transition_probability7}) & (\ref{eq:ste_Pr5}) \\
%\hline
%6 & $m=n=M-1$ & Fig. \ref{fig:m=M-1,n=M-1} & (\ref{eq:transition_probability1}), (\ref{eq:transition_probability3})$-$(\ref{eq:transition_probability8}) & (\ref{eq:ste_Pr6}) \\
%\hline
%7 & $0<m<n<M-1$ & $--$ & $--$ & (\ref{eq:ste_Pr7}) \\
%\hline
%\end{tabular}
%\end{table}

Finally, we summarize the state probabilities and related state transition of the $7$ cases in the second and third columns of Table \ref{tb:vertor_variables}, and the proof is detailed in Appendix A.
%, where Equations (\ref{eq:ste_Pr1})$-$(\ref{eq:ste_Pr7}) are given as follows,

\section{Collision Probability and Throughput Analysis}

%We give the numerical solution to the collision probability and throughput based on the vector form of state probabilities in this section.

\subsection{Numerical Solution of Collision Probability}

For the collision probability in Equation (\ref{eq:collision_pro_eq}),
we adopt Newton method to obtain its numerical solution of $\hat{p}$
since the relationship between $p$ and $\eta$ is highly nonlinear.
After direct calculation, we have the iteration process given by
\begin{equation}
\begin{aligned}
\label{eq:p_updating}
\hat{p}^{(v+1)} = \hat{p}^{(v)} - \bigg[ \Big( 1-\eta|_{p=\hat{p}^{(v)}} \Big)^{N-1} + \hat{p}^{(v)} - 1 \bigg] \bigg[ \Big(N-1\Big)\Big( 1-\eta|_{p=\hat{p}^{(v)}} \Big)^{N-2} \frac{\partial \eta}{\partial p} \bigg|_{p=\hat{p}^{(v)}} \bigg]^{-1},
\end{aligned}
\end{equation}
where $\hat{p}^{(v)}$, $\eta|_{p=\hat{p}^{(v)}}$ and $\frac{\partial \eta}{\partial p}\bigg|_{p=\hat{p}^{(v)}}$ are the numerical values of $p$, $\eta$ and $\frac{\partial \eta}{\partial p}$ in the $v$-th iteration, respectively; and the initial $\hat{p}^{(0)}$ should take value in range $(0,1)$.
For Equation (\ref{eq:p_updating}), we give a method to calculate $\eta$ and $\frac{\partial \eta}{\partial p}$ based on the state transition model.

According to Equation (\ref{eq:eta}), $\eta$ depends on the state probabilities with either $i=0$ or $j=0$.
Let $\epsilon_{m,n} = \mathbb{P}(m,n,0,0), \boldsymbol{\mathrm{r}}_{m,n} = [\mathbb{P}(m,n,0,1), \mathbb{P}(m,n,0,2), \cdots, \mathbb{P}(m,n,0,W_n-1)]^T$ and $\boldsymbol{\mathrm{d}}_{m,n} = [\mathbb{P}(m,n,1,0), \mathbb{P}(m,n,2,0), \cdots, \mathbb{P}(m,n,W_m-1,0)]^T$ for $0 \leq m,n \leq M-1$.
The state probabilities in Equations (\ref{eq:ste_Pr1})-(\ref{eq:ste_Pr7}) can be expressed in vector form in Theorem 1 based on the transition matrices $\boldsymbol{\mathrm{A}}_{r,r,m,n}, \boldsymbol{\mathrm{A}}_{d,r,m,n}, \boldsymbol{\mathrm{A}}_{r,d,m,n}$ and $\boldsymbol{\mathrm{A}}_{d,d,m,n}$ given in Figures \ref{fig:A_rrmn}$-$\ref{fig:A_ddmn}, respectively.

\begin{figure}
\centering
\subfigure[$\boldsymbol{\mathrm{A}}_{r,r,m,n}$.]{
%$\qquad\qquad$
  \begin{minipage}[t]{0.16\linewidth}
    \centering
	\includegraphics[width=1.0\textwidth]{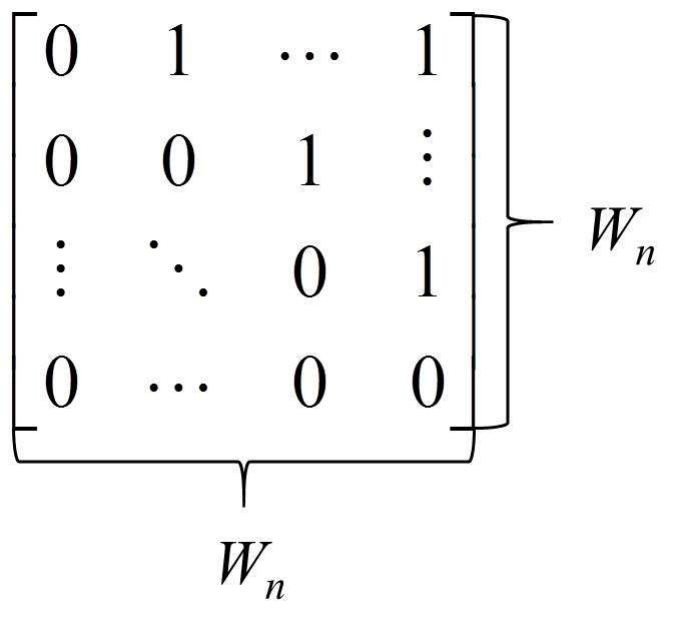}
    \label{fig:A_rrmn}
  \end{minipage}%
}
\subfigure[$\boldsymbol{\mathrm{A}}_{d,r,m,n}$.]{
%$\qquad\qquad\qquad$
  \begin{minipage}[t]{0.26\linewidth}
    \centering
	\includegraphics[width=1.0\textwidth]{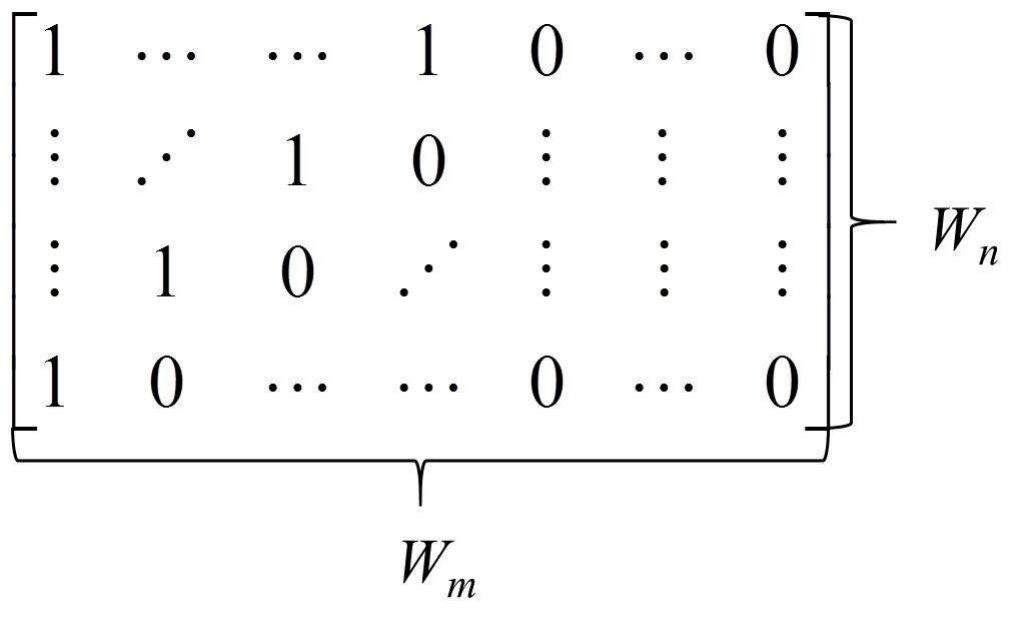}
    \label{fig:A_drmn}
  \end{minipage}%
}
\subfigure[$\boldsymbol{\mathrm{A}}_{r,d,m,n}$.]{
%$\qquad\qquad$
  \begin{minipage}[t]{0.16\linewidth}
    \centering
	\includegraphics[width=1.0\textwidth]{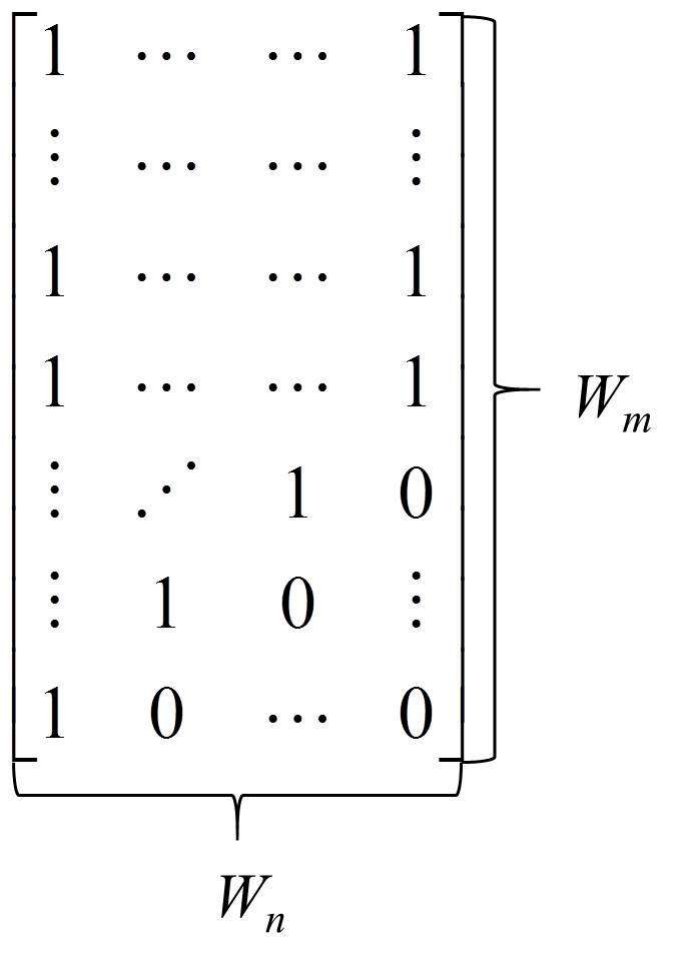}
    \label{fig:A_rdmn}
  \end{minipage}%
}
\subfigure[$\boldsymbol{\mathrm{A}}_{d,d,m,n}$.]{
%$\qquad\qquad\qquad$
  \begin{minipage}[t]{0.26\linewidth}
    \centering
	\includegraphics[width=1.0\textwidth]{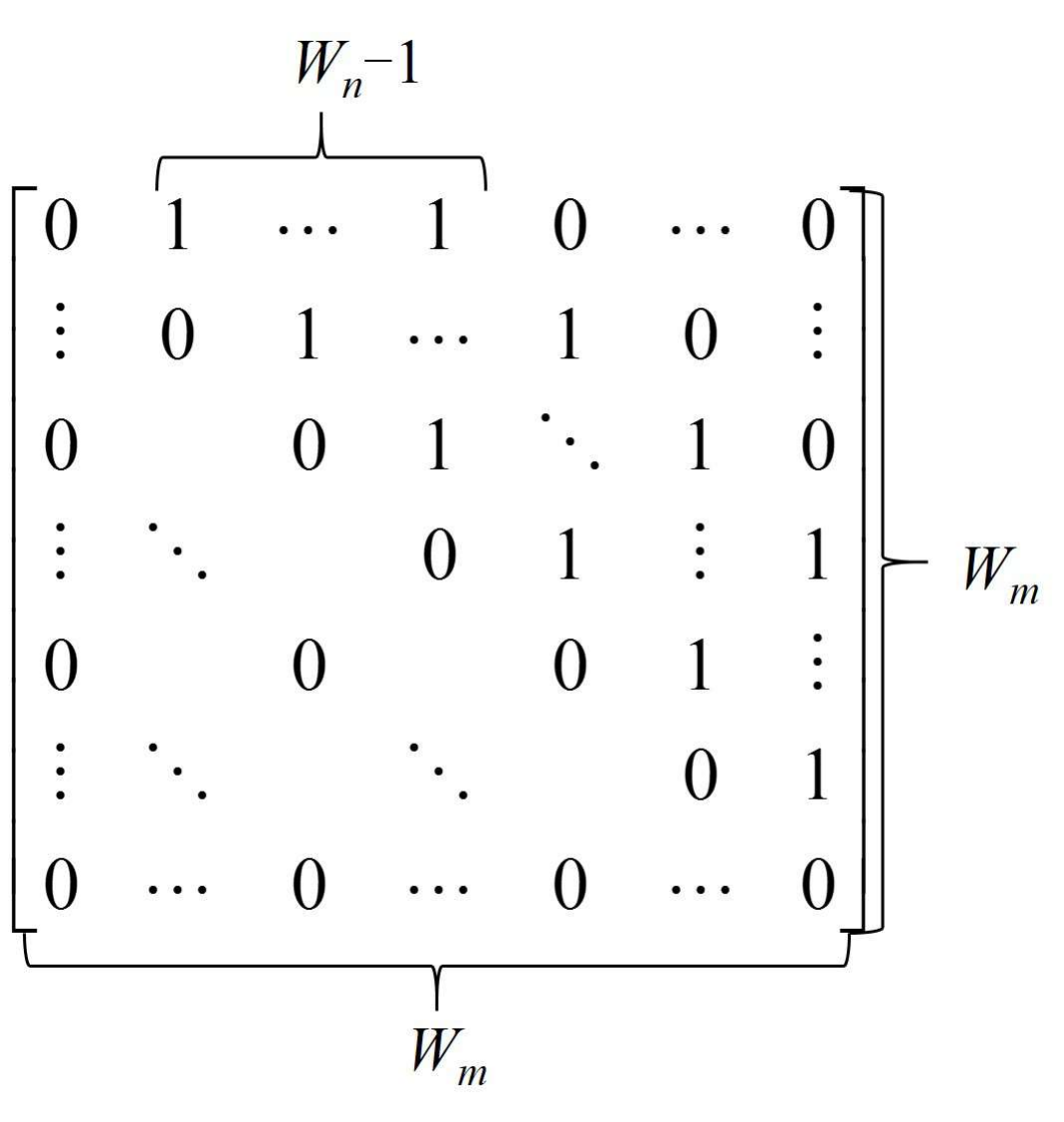}
    \label{fig:A_ddmn}
  \end{minipage}%
}
\caption{The transition matrices.}
\end{figure}

\begin{Theorem}
The vector form of state probabilities corresponding to the $7$ cases in Equations (\ref{eq:ste_Pr1})-(\ref{eq:ste_Pr7}) can be summarized as follows,

\textbf{Case 1}: For $m = 0, n = 0$ and $0 \leq i,j \leq W_0-1$, we have
\begin{equation}
\begin{aligned}
\label{eq:ste_updating1}
r_{0,0,i} = d_{0,0,i} = (W_0-i) W_0^{-1} \epsilon_{0,0}.
\end{aligned}
\end{equation}

\textbf{Case 2}: For $0 < m < M-1$ and $n = 0$, we have
\begin{equation}
\begin{aligned}
\label{eq:ste_updating2}
\boldsymbol{\mathrm{r}}_{m,0} &= W_m^{-1} p \boldsymbol{\mathrm{A}}_{r,r,m,n} \boldsymbol{\mathrm{r}}_{m-1, 0},
\\
\boldsymbol{\mathrm{d}}_{m,0} &= W_m^{-1} p \boldsymbol{\mathrm{A}}_{r,d,m,n} \boldsymbol{\mathrm{r}}_{m-1, 0},
\\
\epsilon_{m,0} &= W_m^{-1} p \boldsymbol{1}^T_{W_n-1} \boldsymbol{\mathrm{r}}_{m-1, 0}.
\end{aligned}
\end{equation}

\textbf{Case 3}: For $n = 0$ and $m = M-1$, we have
\begin{equation}
\begin{aligned}
\label{eq:ste_updating3}
\boldsymbol{\mathrm{r}}_{M-1,0} &= (\boldsymbol{I} - W_{M-1}^{-1} p \boldsymbol{\mathrm{A}}_{r,r,m,n})^{-1} ( W_{M-1}^{-1} p \boldsymbol{\mathrm{A}}_{r,r,m,n} \boldsymbol{\mathrm{r}}_{M-2, 0} ),
%\boldsymbol{\mathrm{r}}_{M-1,0} &= W_{M-1}^{-1} p \boldsymbol{\mathrm{A}}_{r,r,m,n} (\boldsymbol{\mathrm{r}}_{M-2, 0} + \boldsymbol{\mathrm{r}}_{M-1, 0}),
\\
\boldsymbol{\mathrm{d}}_{M-1,0} &= W_{M-1}^{-1} p \boldsymbol{\mathrm{A}}_{r,d,m,n} (\boldsymbol{\mathrm{r}}_{M-2, 0} + \boldsymbol{\mathrm{r}}_{M-1, 0}),
\\
\epsilon_{M-1,0} &= W_{M-1}^{-1} p \boldsymbol{1}^T_{W_n-1} (\boldsymbol{\mathrm{r}}_{M-2, 0} + \boldsymbol{\mathrm{r}}_{M-1, 0}).
\end{aligned}
\end{equation}
%where $\boldsymbol{\mathrm{r}}_{M-1,0}$ can be simplified as
%\begin{equation}
%\begin{aligned}
%\boldsymbol{\mathrm{r}}_{M-1,0} &= (\boldsymbol{I} - W_{M-1}^{-1} p \boldsymbol{\mathrm{A}}_{r,r,m,n})^{-1} ( W_{M-1}^{-1} p \boldsymbol{\mathrm{A}}_{r,r,m,n} \boldsymbol{\mathrm{r}}_{M-2, 0} ).
%\end{aligned}
%\end{equation}

\textbf{Case 4}: For $0 < n \leq m < M-1$, we have
\begin{equation}
\begin{aligned}
\label{eq:ste_updating4}
\boldsymbol{\mathrm{r}}_{m,n} &= W_m^{-1} p \boldsymbol{\mathrm{A}}_{r,r,m,n} \boldsymbol{\mathrm{r}}_{m-1, n} + W_n^{-1} p \boldsymbol{\mathrm{A}}_{d,r,m,n} \boldsymbol{\mathrm{d}}_{m, n-1} + W_m^{-1} W_n^{-1} \epsilon_{m-1,n-1} \boldsymbol{\mathrm{u}}_{n,n},
\\
\boldsymbol{\mathrm{d}}_{m,n} &= W_m^{-1} p \boldsymbol{\mathrm{A}}_{r,d,m,n} \boldsymbol{\mathrm{r}}_{m-1, 0} + W_n^{-1} p \boldsymbol{\mathrm{A}}_{d,d,m,n} \boldsymbol{\mathrm{d}}_{m, n-1} + W_m^{-1} W_n^{-1} \epsilon_{m-1,n-1} \boldsymbol{\mathrm{u}}_{m,n},
\\
\epsilon_{m,n} &= W_m^{-1} p \boldsymbol{1}^T_{W_n-1} \boldsymbol{\mathrm{r}}_{m-1, n} + W_n^{-1} p \boldsymbol{1}^T_{W_n} \boldsymbol{\mathrm{d}}_{m, n-1, [1,W_n]} + W_m^{-1} \epsilon_{m-1,n-1},
\end{aligned}
\end{equation}
where $\boldsymbol{u}_{m,n} = \underbrace{[W_n, W_n, \cdots, W_n, W_n-1, \cdots, 1]}_{W_m}$.

\textbf{Case 5}: For $m = M-1$ and $0 < n < M-1$, we have
\begin{equation}
\begin{aligned}
\label{eq:ste_updating5}
\boldsymbol{\mathrm{r}}_{M-1,n} \!&=\! (\boldsymbol{I} \!\!-\!\! W_{M-1}^{-1} \! p \boldsymbol{\mathrm{A}}_{r,r,m,n})^{-1} \! \big[ W_{M-1}^{-1} \!p \boldsymbol{\mathrm{A}}_{r,r,m,n} \boldsymbol{\mathrm{r}}_{M-2, n} \!\!+\!\!  W_n^{-1} \!p \boldsymbol{\mathrm{A}}_{d,r,m,n} \boldsymbol{\mathrm{d}}_{M-1, n-1} \!\!+\!\! W_{M-1}^{-1} \!\! W_n^{-1} \! ( \! \epsilon_{M-2,n-1} \!\!+\!  \epsilon_{M-1,n-1} \!) \boldsymbol{\mathrm{u}}_{n,n} \big],
%\boldsymbol{\mathrm{r}}_{M-1,n} &= W_{M-1}^{-1} p \boldsymbol{\mathrm{A}}_{r,r,m,n} ( \boldsymbol{\mathrm{r}}_{M-2, n} + \boldsymbol{\mathrm{r}}_{M-1, n}) + W_n^{-1} p \boldsymbol{\mathrm{A}}_{d,r,m,n} \boldsymbol{\mathrm{d}}_{M-1, n-1} + W_{M-1}^{-1} W_n^{-1} (\epsilon_{M-2,n-1} + \epsilon_{M-1,n-1}) \boldsymbol{\mathrm{u}}_{n,n},
\\
\boldsymbol{\mathrm{d}}_{M-1,n} &= W_{M-1}^{-1} p \boldsymbol{\mathrm{A}}_{r,d,m,n} ( \boldsymbol{\mathrm{r}}_{M-2, n} \!+\! \boldsymbol{\mathrm{r}}_{M-1, n}) \!+\! W_n^{-1} p \boldsymbol{\mathrm{A}}_{d,d,m,n} \boldsymbol{\mathrm{d}}_{M-1, n-1} \!+\! W_{M-1}^{-1} W_n^{-1} (\epsilon_{M-2,n-1} \!+\! \epsilon_{M-1,n-1}) \boldsymbol{\mathrm{u}}_{M-1,n},
\\
\epsilon_{M-1,n} &= W_{M-1}^{-1} p \boldsymbol{1}^T_{W_n-1} ( \boldsymbol{\mathrm{r}}_{M-2, n} + \boldsymbol{\mathrm{r}}_{M-1, n}) + W_n^{-1} p \boldsymbol{1}^T_{W_n} \boldsymbol{\mathrm{d}}_{M-1, n-1, [1,W_n]} + W_{M-1}^{-1} (\epsilon_{M-2,n-1} + \epsilon_{M-1,n-1}).
\end{aligned}
\end{equation}
%where $\boldsymbol{\mathrm{r}}_{M-1,n}$ can be determined by
%\begin{equation}
%\begin{aligned}
%\boldsymbol{\mathrm{r}}_{M-1,n} &= (\boldsymbol{I} - W_{M-1}^{-1} p \boldsymbol{\mathrm{A}}_{r,r,m,n})^{-1} ( W_{M-1}^{-1} p \boldsymbol{\mathrm{A}}_{r,r,m,n} \boldsymbol{\mathrm{r}}_{M-2, n} + W_n^{-1} p \boldsymbol{\mathrm{A}}_{d,r,m,n} \boldsymbol{\mathrm{d}}_{M-1, n-1} + W_{M-1}^{-1} W_n^{-1} \epsilon^{\sigma} \boldsymbol{\mathrm{u}}_{n,n} );
%\end{aligned}
%\end{equation}
%and
%where $\epsilon^{\sigma} = \epsilon_{M-2,n-1} + \epsilon_{M-1,n-1}$.

\textbf{Case 6}: For $m = n = M-1$, we have
%\begin{equation}
%\begin{aligned}
%\label{eq:ste_updating6}
%\boldsymbol{\mathrm{r}}_{M-1,M-1} &= W_{M-1}^{-1} p \boldsymbol{\mathrm{A}}_{r,r,M-1,M-1} ( \boldsymbol{\mathrm{r}}_{M-2, M-1} + \boldsymbol{\mathrm{r}}_{M-1, M-1}) + W_{M-1}^{-1} p \boldsymbol{\mathrm{A}}_{d,r,M-1,M-1} (\boldsymbol{\mathrm{d}}_{M-1, M-2} + \boldsymbol{\mathrm{d}}_{M-1, M-1})
%\\
%&+ W_{M-1}^{-2} (\epsilon_{M-2,M-2} + \epsilon_{M-2,M-1} + \epsilon_{M-1,M-2} + \epsilon_{M-1,M-1}) \boldsymbol{\mathrm{u}}_{M-1,M-1},
%\\
%\boldsymbol{\mathrm{d}}_{M-1,M-1} &= \boldsymbol{\mathrm{r}}_{M-1,M-1},
%\\
%\epsilon_{M-1,M-1} &= W_{M-1}^{-1} p \boldsymbol{1}^T_{W_{M-1}-1} ( \boldsymbol{\mathrm{r}}_{M-2, M-1} + \boldsymbol{\mathrm{r}}_{M-1, M-1}) + W_{M-1}^{-1} p \boldsymbol{1}^T_{W_{M-1}-1} (\boldsymbol{\mathrm{d}}_{M-1, M-2} + \boldsymbol{\mathrm{d}}_{M-1, M-1})
%\\
%&+ W_{M-1}^{-1} (\epsilon_{M-2,M-2} + \epsilon_{M-2,M-1} + \epsilon_{M-1,M-2} + \epsilon_{M-1,M-1}),
%\end{aligned}
%\end{equation}
%and the simplified expression is given as follows,
\begin{equation}
\begin{aligned}
\label{eq:ste_updating6}
\epsilon_{M-1,M-1} &= \Big[ 1 - 2 W_{M-1}^{-3} p \boldsymbol{1}^T_{W_{M-1}-1} \big( \boldsymbol{I} - W_{M-1}^{-1} p \boldsymbol{A}^{\sigma} \big)^{-1} \boldsymbol{u}_{M-1,M-1} - W_{M-1}^{-1} \Big]^{-1}
\Big[ 2 W_{M-1}^{-1} p \boldsymbol{1}^T_{W_{M-1}-1} \boldsymbol{\mathrm{d}}_{M-1, M-2}
\\
&+ W_{M-1}^{-1} \epsilon^{\sigma} + 2 W_{M-1}^{-1} p \boldsymbol{1}^T_{W_{M-1}-1} \big( \boldsymbol{I} - W_{M-1}^{-1} p \boldsymbol{A}^{\sigma} \big)^{-1} \big( W_{M-1}^{-1} p \boldsymbol{A}^{\sigma} \boldsymbol{\mathrm{d}}_{M-1, M-2} + W_{M-1}^{-2} \epsilon^{\sigma} \boldsymbol{u}_{M-1,M-1} \big) \Big],
\\
\boldsymbol{\mathrm{r}}_{M-1,M-1} &= \boldsymbol{\mathrm{d}}_{M-1,M-1}
\\
&= \big( \boldsymbol{I} - W_{M-1}^{-1} p \boldsymbol{A}^{\sigma} \big)^{-1} \big[ W_{M-1}^{-1} p \boldsymbol{A}^{\sigma} \boldsymbol{\mathrm{d}}_{M-1, M-2} + W_{M-1}^{-2} ( \epsilon^{\sigma} + \epsilon_{M-1,M-1}) \boldsymbol{u}_{M-1,M-1} \big],
\end{aligned}
\end{equation}
where $\boldsymbol{A}^{\sigma} = \boldsymbol{A}_{r,r,M-1,M-1} + \boldsymbol{A}_{r,d,M-1,M-1} = \boldsymbol{A}_{d,d,M-1,M-1} + \boldsymbol{A}_{d,r,M-1,M-1}$ and $\epsilon^{\sigma} = \epsilon_{M-2,M-2} + \epsilon_{M-2,M-1} + \epsilon_{M-1,M-2}$.

\textbf{Case 7}: For $0 \leq m < n \leq M-1$, we have $\boldsymbol{\mathrm{r}}_{m,n} = \boldsymbol{\mathrm{d}}_{n,m}, \boldsymbol{\mathrm{d}}_{m,n} = \boldsymbol{\mathrm{r}}_{n,m}, \epsilon_{m,n} = \epsilon_{n,m}$.
%\begin{equation}
%\begin{aligned}
%\boldsymbol{\mathrm{r}}_{m,n} &= \boldsymbol{\mathrm{d}}_{n,m}
%\\
%\boldsymbol{\mathrm{d}}_{m,n} &= \boldsymbol{\mathrm{r}}_{n,m}
%\\
%\epsilon_{m,n} &= \epsilon_{n,m}
%\end{aligned}
%\end{equation}
\end{Theorem}

\emph{Proof:} Please refer to Appendix B.

Based on Theorem 1, we can use $\epsilon_{0,0}$ to obtain $r_{m,n,i}, d_{m,n,j}$ and $\epsilon_{m,n}$ for $0 \leq m,n \leq M-1$, where $r_{m,n,i}$ and $d_{m,n,j}$ are the $i$-th and $j$-th element of $\boldsymbol{\mathrm{r}}_{m,n}$ and $\boldsymbol{\mathrm{d}}_{m,n}$, respectively, for $0 \leq i \leq W_m - 1, 0 \leq j \leq W_n - 1$.
%Furthermore, it is convinced that $\sum_{m=0}^{M-1} \sum_{n=0}^{M-1} \sum_{i=0}^{W_m-1} \sum_{j=0}^{W_n-1} \mathbb{P}(m,n,i,j) = 1$, which means that the summation of all probabilities equals $1$.
Since $\sum_{m=0}^{M-1} \sum_{n=0}^{M-1} \sum_{i=0}^{W_m-1} \sum_{j=0}^{W_n-1} \mathbb{P}(m,n,i,j) = 1$, letting $\mathbb{P}(m,n) = \sum_{i=0}^{W_m-1} \sum_{j=0}^{W_n-1} \mathbb{P}(m,n,i,j)$ denote the probability that the contention windows of $T_1$ and $T_2$ equal $W_m$ and $W_n$, respectively, we have the following Theorem 2 to calculate $\mathbb{P}(m,n)$ for $0 \leq m,n \leq M-1$.

% Hence, for calculating $\sum_{m=0}^{M-1} \sum_{n=0}^{M-1} \sum_{i=0}^{W_m-1} \sum_{j=0}^{W_n-1} \mathbb{P}(m,n,i,j)$, we achieve $\mathbb{P}(m,n) = \sum_{i=0}^{W_m-1} \sum_{j=0}^{W_n-1} \mathbb{P} (m,n,i,j)$ based on Theorem 2, where $\mathbb{P}(m,n)$ denotes the probability that the contention windows of $T_1$ and $T_2$ equal $W_m$ and $W_n$, respectively.

\begin{Theorem}
Corresponding to the $7$ cases in Theorem 1, probability $\mathbb{P}(m,n)$ can be calculated as follows,

\textbf{Case 1}: For $m = n = 0$, we have
\begin{equation}
\begin{aligned}
\label{eq:sigma_r_d1}
\mathbb{P}(0,0) = \frac{1}{6} (2 W_0 + 1) (W_0 + 1) \epsilon_{0,0}. % \bigg( \frac{1}{3} W_0^3 + \frac{1}{2} W_0^2 + \frac{1}{6} W_0 \bigg)\epsilon_{0,0}. %
\end{aligned}
\end{equation}%\frac{1}{3} W_0^3 + \frac{1}{2} W_0^2 + \frac{1}{6} W_0$.

\textbf{Case 2}: For $0 < m < M-1$ and $n = 0$, we have
\begin{equation}
\begin{aligned}
\label{eq:sigma_r_d2}
\mathbb{P}(m,0) = W_m^{-1} p \sum^{W_0-1}_{i=1} r_{m-1,0,i} \bigg[ -\frac{1}{2} i^2 + \Big(W_m + \frac{1}{2}\Big)i \bigg].
\end{aligned}
\end{equation}

\textbf{Case 3}: For $m = M-1$ and $n = 0$, we have
\begin{equation}
\begin{aligned}
\label{eq:sigma_r_d3}
\mathbb{P}(M-1,0) = W_{M-1}^{-1} p \sum^{W_0-1}_{i=1} (r_{M-2,0,i} + r_{M-1,0,i}) \bigg[ -\frac{1}{2} i^2 + \Big(W_{M-1} + \frac{1}{2}\Big)i \bigg].
\end{aligned}
\end{equation}

\textbf{Case 4}: For $0 < n \leq m < M-1$, we have
\begin{equation}
\begin{aligned}
\label{eq:sigma_r_d4}
\mathbb{P}(m,n) &= W_m^{-1} p \sum^{W_n-1}_{i=1} r_{m-1,n,i} \bigg[ -\frac{1}{2} i^2 + \Big(W_m + \frac{1}{2}\Big)i \bigg] + W_n^{-1} p \sum^{W_n-1}_{i=1} d_{m,n-1,i} \bigg[ -\frac{1}{2} i^2 + \Big(W_n + \frac{1}{2}\Big)i \bigg]
\\
&+\! W_n^{-1} \! p \!\! \sum^{W_m-1}_{i=W_n} \!\! d_{m,n-1,i} \Big( \!-\! \frac{1}{2} W_n^2 \!+\! \frac{1}{2} W_n \Big) \!+\! W_m^{-1} W_n^{-1} \epsilon_{m-1,n-1} \Big( \!-\! \frac{1}{6}W_n^3 \!+\! \frac{1}{2} W_m W_n^2 \!+\! \frac{1}{2} W_m W_n \!+\! \frac{1}{6} W_n \Big).
\end{aligned}
\end{equation}

\textbf{Case 5}: For $m = M-1$ and $0 < n < M-1$, we have
\begin{equation}
\begin{aligned}
\label{eq:sigma_r_d5}
\mathbb{P}(M-1,n) &= W_{M-1}^{-1} p \!\!\!\! \sum^{W_n-1}_{i=1} \!\! (r_{M-2,n,i} \!+\! r_{M-1,n,i}) \bigg[ -\frac{1}{2} i^2 \!+\! \Big(W_{M-1} \!+\! \frac{1}{2}\Big)i \bigg] \!+\! W_n^{-1} p \!\!\!\! \sum^{W_n-1}_{i=1} \!\!\!\! d_{M-1,n-1,i} \bigg[ -\frac{1}{2} i^2 \!+\! \Big(W_n \!+\! \frac{1}{2}\Big)i \bigg]
\\
&+\!\! W_n^{-1} p \!\!\!\!\! \sum^{W_{M-1}-1}_{i=W_n} \!\!\!\!\! d_{M-1,n-1,i} \Big( -\!\! \frac{1}{2} W_n^2 \!+\! \frac{1}{2} W_n \Big) \!+\! W_{M-1}^{-1} W_n^{-1} \epsilon^{\sigma} \Big( -\!\! \frac{1}{6}W_n^3 \!\!+\!\! \frac{1}{2} W_{M-1} W_n^2 \!\!+\!\! \frac{1}{2} W_{M-1} W_n \!\!+\!\! \frac{1}{6} W_n \Big),
\end{aligned}
\end{equation}
where $\epsilon^{\sigma} = \epsilon_{M-2,n-1} + \epsilon_{M-1,n-1}$.

\textbf{Case 6}: For $m = M-1$ and $n = M-1$, we have
\begin{equation}
\begin{aligned}
\label{eq:sigma_r_d6}
\mathbb{P}(M-1,M-1) &= W_{M-1}^{-1} p \sum^{W-1}_{i=1} d^{\sigma}_i \bigg[ -\frac{1}{2} i^2 + \Big(W_{M-1} + \frac{1}{2}\Big)i \bigg] + W_{M-1}^{-2} \epsilon^{\sigma} \Big(\frac{1}{3}W_{M-1}^3 + \frac{1}{2}W_{M-1}^2 + \frac{1}{6}W_{M-1} \Big),
\end{aligned}
\end{equation}
where $d^{\sigma}_i = r_{m-1,n,i} + r_{m,n,i} + d_{m,n-1,i} + d_{m,n,i}$ and $\epsilon^{\sigma} = \epsilon_{M-2,M-2} + \epsilon_{M-2,M-1} + \epsilon_{M-1,M-2} + \epsilon_{M-1,M-1}$.

\textbf{Case 7}: For $0 \leq m < n \leq M-1$, we have that $\mathbb{P}(m,n) = \mathbb{P}(n,m)$.

\end{Theorem}

\emph{Proof:} Please refer to Appendix C.

Based on Theorems 1 and 2, we can utilize $\epsilon_{0,0}$ to obtain $\mathbb{P}(m,n)$ for $0 \leq m,n \leq M-1$, and adopt the condition that $\sum_{m=0}^{M-1} \sum_{n=0}^{M-1} \mathbb{P}(m,n) = 1$ to calculate $\epsilon_{0,0}$.
Furtherly, we adopt Theorem 1 to obtain $r_{m,n,i}, d_{m,n,j}$ and $\epsilon_{m,n}$ for $0 \leq i \leq W_m - 1, 0 \leq j \leq W_n - 1, 0 \leq m,n \leq M-1$ based on $\epsilon_{0,0}$, and calculate $\eta |_{p = \hat{p}^{(v)}}$ according to Equation (\ref{eq:eta}) or its equivalent vector as follows,
\begin{equation}
\begin{aligned}
\label{eq:eta2}
\eta =
\sum^{M-1}_{m=0} \sum^{M-1}_{n=0} \boldsymbol{1}^T_{W_n-1} \boldsymbol{\mathrm{r}}_{m,n} + \boldsymbol{1}^T_{W_m-1} \boldsymbol{\mathrm{d}}_{m,n}.
\end{aligned}
\end{equation}

Moreover, in order to calculate $\frac{\partial \eta}{\partial p}|_{p = \hat{p}^{(v)}}$ in Equation (\ref{eq:p_updating}), we have that
\begin{equation}
\begin{aligned}
\label{eq:eta_derivate_p}
\frac{\partial \eta}{\partial p} = \sum_{m=0}^{M-1} \sum_{n=0}^{M-1} \boldsymbol{1}^T_{W_n-1} \nabla_p \boldsymbol{\mathrm{r}}_{m,n} + \boldsymbol{1}^T_{W_m-1} \nabla_p \boldsymbol{\mathrm{d}}_{m,n},
\end{aligned}
\end{equation}
where $\nabla_p \boldsymbol{\mathrm{r}}_{m,n} = \Big[ \frac{\partial \mathbb{P}(m,n,0,1)}{\partial p}, \frac{\partial \mathbb{P}(m,n,0,2)}{\partial p}, \cdots, \frac{\partial \mathbb{P}(m,n,0,\mathbb{P}(m,n,0,W_n-1))}{\partial p} \Big]^T$ and $\nabla_p \boldsymbol{\mathrm{d}}_{m,n} = \Big[ \frac{\partial \mathbb{P}(m,n,1,0)}{\partial p}, \frac{\partial \mathbb{P}(m,n,2,0)}{\partial p}, \\ \cdots, \frac{\partial \mathbb{P}(m,n,W_m-1,0)}{\partial p} \Big]^T$.
%\textcolor{red}{It is persuasive that $\frac{\partial \eta}{\partial p}$ is a special case of Equation (\ref{eq:eta_derivate_p}) where $k = 1$, and the case that $k = 2$ can be applied to calculate $\frac{\partial^2 \eta}{\partial p^2}$ for Appendix F.}
According to Theorem 1, we have Lemma 1 to calculate $\nabla_p \boldsymbol{\mathrm{r}}_{m,n}, \nabla_p \boldsymbol{\mathrm{d}}_{m,n}$ and $\frac{\partial \epsilon_{m,n}}{\partial p}$ as follows, where the proof is omitted since it is based on standard calculus.

\begin{Lemma}
Corresponding to the $7$ cases in Theorem 1, $\nabla_p \boldsymbol{\mathrm{r}}_{m,n}, \nabla_p \boldsymbol{\mathrm{d}}_{m,n}$ and $\frac{\partial \epsilon_{m,n}}{\partial p}$ can be characterized as follows,

\textbf{Case 1}: For $m = n = 0$, and $0 \leq i,j \leq W_0-1$, we have
\begin{equation}
\begin{aligned}
\label{eq:dste_updating1}
\frac{\partial r_{0,0,i}}{\partial p} = \frac{\partial d_{0,0,i}}{\partial p} = (W_0-i) W_0^{-1} \frac{\partial \epsilon_{0,0}}{\partial p}.
\end{aligned}
\end{equation}

\textbf{Case 2}: For $0 < m < M-1$ and $n = 0$, we have
\begin{equation}
\begin{aligned}
\label{eq:dste_updating2}
\nabla_p \boldsymbol{\mathrm{r}}_{m,0} &= W_m^{-1} \boldsymbol{\mathrm{A}}_{r,r,m,n} ( \nabla_p \boldsymbol{\mathrm{r}}_{m-1, 0} + p \nabla_p \boldsymbol{\mathrm{r}}_{m-1, 0} ),
\\
\nabla_p \boldsymbol{\mathrm{d}}_{m,0} &= W_m^{-1} \boldsymbol{\mathrm{A}}_{r,d,m,n} (  \nabla_p \boldsymbol{\mathrm{r}}_{m-1, 0} + p \nabla_p \boldsymbol{\mathrm{r}}_{m-1, 0} ),
\\
\frac{\partial \epsilon_{m,0}}{\partial p} &= W_m^{-1} \boldsymbol{1}^T_{W_n-1} ( \nabla_p \boldsymbol{\mathrm{r}}_{m-1, 0} + p \nabla_p \boldsymbol{\mathrm{r}}_{m-1, 0} ).
\end{aligned}
\end{equation}

\textbf{Case 3}: For $m = M-1$ and $n = 0$, we have
\begin{equation}
\begin{aligned}
\label{eq:dste_updating3}
\nabla_p \boldsymbol{\mathrm{r}}_{M-1,0} &= (\boldsymbol{I} - W_{M-1}^{-1} p \boldsymbol{\mathrm{A}}_{r,r,m,n})^{-1} W_{M-1}^{-1} p \boldsymbol{\mathrm{A}}_{r,r,m,n} (\nabla_p \boldsymbol{\mathrm{r}}_{M-2, 0} + \nabla_p \boldsymbol{\mathrm{r}}_{M-1, 0} + p \nabla_p \boldsymbol{\mathrm{r}}_{M-2, 0}),
%\nabla_p \boldsymbol{\mathrm{r}}_{M-1,0} &= W_{M-1}^{-1} \boldsymbol{\mathrm{A}}_{r,r,m,n} (\nabla_p \boldsymbol{\mathrm{r}}_{M-2, 0} + \nabla_p \boldsymbol{\mathrm{r}}_{M-1, 0} + p \nabla_p \boldsymbol{\mathrm{r}}_{M-2, 0} + p \nabla_p \boldsymbol{\mathrm{r}}_{M-1, 0}),
\\
\nabla_p \boldsymbol{\mathrm{d}}_{M-1,0} &= W_{M-1}^{-1} \boldsymbol{\mathrm{A}}_{r,d,m,n} (\nabla_p \boldsymbol{\mathrm{r}}_{M-2, 0} + \nabla_p \boldsymbol{\mathrm{r}}_{M-1, 0} + p \nabla_p \boldsymbol{\mathrm{r}}_{M-2, 0} + p \nabla_p \boldsymbol{\mathrm{r}}_{M-1, 0}),
\\
\frac{\partial \epsilon_{M-1,0}}{\partial p} &= W_{M-1}^{-1} \boldsymbol{1}^T_{W_n-1} (\nabla_p \boldsymbol{\mathrm{r}}_{M-2, 0} + \nabla_p \boldsymbol{\mathrm{r}}_{M-1, 0} + p \nabla_p \boldsymbol{\mathrm{r}}_{M-2, 0} + p \nabla_p \boldsymbol{\mathrm{r}}_{M-1, 0}).
\end{aligned}
\end{equation}
%where $\nabla_p \boldsymbol{\mathrm{r}}_{M-1,0}$ can be simplified into
%\begin{equation}
%\begin{aligned}
%\nabla_p \boldsymbol{\mathrm{r}}_{M-1,0} &= (\boldsymbol{I} - W_{M-1}^{-1} p \boldsymbol{\mathrm{A}}_{r,r,m,n})^{-1} W_{M-1}^{-1} p \boldsymbol{\mathrm{A}}_{r,r,m,n} (\nabla_p \boldsymbol{\mathrm{r}}_{M-2, 0} + \nabla_p \boldsymbol{\mathrm{r}}_{M-1, 0} + p \nabla_p \boldsymbol{\mathrm{r}}_{M-2, 0})
%\end{aligned}
%\end{equation}

\textbf{Case 4}: For $0 < n \leq m < M-1$, we have
\begin{equation}
\begin{aligned}
\label{eq:dste_updating4}
\nabla_p \boldsymbol{\mathrm{r}}_{m,n} &= W_m^{-1} \boldsymbol{\mathrm{A}}_{r,r,m,n} (\nabla_p \boldsymbol{\mathrm{r}}_{m-1, n} \!\!+\!\! p \nabla_p \boldsymbol{\mathrm{r}}_{m-1, n}) \!\!+\!\! W_n^{-1} \boldsymbol{\mathrm{A}}_{d,r,m,n} (\nabla_p \boldsymbol{\mathrm{d}}_{m, n-1} \!\!+\!\! p \nabla_p \boldsymbol{\mathrm{d}}_{m, n-1}) \!\!+\!\! W_m^{-1} W_n^{-1} \frac{\partial \epsilon_{m-1,n-1}}{\partial p} \boldsymbol{\mathrm{u}}_{n,n},
\\
\nabla_p \boldsymbol{\mathrm{d}}_{m,n} &= W_m^{-1} \boldsymbol{\mathrm{A}}_{r,d,m,n} (\nabla_p \boldsymbol{\mathrm{r}}_{m-1, n} \!\!+\!\! p \nabla_p \boldsymbol{\mathrm{r}}_{m-1, n}) \!\!+\!\! W_n^{-1} \boldsymbol{\mathrm{A}}_{d,d,m,n} (\nabla_p \boldsymbol{\mathrm{d}}_{m, n-1} \!\!+\!\! p \nabla_p \boldsymbol{\mathrm{d}}_{m, n-1}) \!\!+\!\! W_m^{-1} W_n^{-1} \frac{\partial \epsilon_{m-1,n-1}}{\partial p} \boldsymbol{\mathrm{u}}_{m,n},
\\
\frac{\partial \epsilon_{m,n}}{\partial p}  &= W_m^{-1}  \boldsymbol{1}^T_{W_n-1} (\nabla_p \boldsymbol{\mathrm{r}}_{m-1, n} \!+\! p \nabla_p \boldsymbol{\mathrm{r}}_{m-1, n}) \!+\! W_n^{-1} \boldsymbol{1}^T_{W_n} (\nabla_p \boldsymbol{\mathrm{d}}_{m, n-1, [1,W_n]} \!+\! p \nabla_p \boldsymbol{\mathrm{d}}_{m, n-1, [1,W_n]}) \!+\! W_m^{-1} \frac{\partial \epsilon_{m-1,n-1}}{\partial p}.
\end{aligned}
\end{equation}

\textbf{Case 5}: For $m = M-1$ and $0 < n < M-1$, we have
\begin{equation}
\begin{aligned}
\label{eq:dste_updating5}
\nabla_p \boldsymbol{\mathrm{r}}_{M-1,n} &= (\boldsymbol{I} - W_{M-1}^{-1} p \boldsymbol{\mathrm{A}}_{r,r,m,n})^{-1} \bigg[ W_{M-1}^{-1} \boldsymbol{\mathrm{A}}_{r,r,m,n} ( \nabla_p \boldsymbol{\mathrm{r}}_{M-2, n} + \nabla_p \boldsymbol{\mathrm{r}}_{M-1, n} + p \nabla_p \boldsymbol{\mathrm{r}}_{M-2, n} ) + W_n^{-1} \boldsymbol{\mathrm{A}}_{d,r,m,n}
\\
&( \nabla_p \boldsymbol{\mathrm{d}}_{M-1, n-1} + p \nabla_p \boldsymbol{\mathrm{d}}_{M-1, n-1}) + W_{M-1}^{-1} W_n^{-1} \Big( \frac{\partial \epsilon_{M-2,n-1}}{\partial p} + \frac{\partial \epsilon_{M-1,n-1}}{\partial p} \Big) \boldsymbol{\mathrm{u}}_{n,n} \bigg],
%\nabla_p \boldsymbol{\mathrm{r}}_{M-1,n} &= W_{M-1}^{-1} \boldsymbol{\mathrm{A}}_{r,r,m,n} ( \nabla_p \boldsymbol{\mathrm{r}}_{M-2, n} + \nabla_p \boldsymbol{\mathrm{r}}_{M-1, n} + p \nabla_p \boldsymbol{\mathrm{r}}_{M-2, n} + p \nabla_p \boldsymbol{\mathrm{r}}_{M-1, n} ) + W_n^{-1} \boldsymbol{\mathrm{A}}_{d,r,m,n} ( \nabla_p \boldsymbol{\mathrm{d}}_{M-1, n-1} +
%\\
%&p \nabla_p \boldsymbol{\mathrm{d}}_{M-1, n-1}) + W_{M-1}^{-1} W_n^{-1} \Big( \frac{\partial \epsilon_{M-2,n-1}}{\partial p} + \frac{\partial \epsilon_{M-1,n-1}}{\partial p} \Big) \boldsymbol{\mathrm{u}}_{n,n}
\\
\nabla_p \boldsymbol{\mathrm{d}}_{M-1,n} &= W_{M-1}^{-1} \boldsymbol{\mathrm{A}}_{r,d,m,n} ( \nabla_p \boldsymbol{\mathrm{r}}_{M-2, n} + \nabla_p \boldsymbol{\mathrm{r}}_{M-1, n} + p \nabla_p \boldsymbol{\mathrm{r}}_{M-2, n} + p \nabla_p \boldsymbol{\mathrm{r}}_{M-1, n} ) + W_n^{-1} p \boldsymbol{\mathrm{A}}_{d,d,m,n} ( \nabla_p \boldsymbol{\mathrm{d}}_{M-1, n-1} +
\\
&p \nabla_p \boldsymbol{\mathrm{d}}_{M-1, n-1}) + W_{M-1}^{-1} W_n^{-1} \Big( \frac{\partial \epsilon_{M-2,n-1}}{\partial p} + \frac{\partial \epsilon_{M-1,n-1}}{\partial p} \Big) \boldsymbol{\mathrm{u}}_{M-1,n},
\\
\frac{\partial \epsilon_{M-1,n}}{\partial p} &= W_{M-1}^{-1} \boldsymbol{1}^T_{W_n-1} ( \nabla_p \boldsymbol{\mathrm{r}}_{M-2, n} + \nabla_p \boldsymbol{\mathrm{r}}_{M-1, n} + p \nabla_p \boldsymbol{\mathrm{r}}_{M-2, n} + p \nabla_p \boldsymbol{\mathrm{r}}_{M-1, n} ) + W_n^{-1} \boldsymbol{1}^T_{W_n} ( \nabla_p \boldsymbol{\mathrm{d}}_{M-1, n-1, [1,W_n]} +
\\
&p \nabla_p \boldsymbol{\mathrm{d}}_{M-1, n-1, [1,W_n]}) + W_{M-1}^{-1} \Big( \frac{\partial \epsilon_{M-2,n-1}}{\partial p} + \frac{\partial \epsilon_{M-1,n-1}}{\partial p} \Big).
\end{aligned}
\end{equation}
%where $\boldsymbol{\mathrm{r}}^{\sigma} = \boldsymbol{\mathrm{r}}_{M-2, n} + \boldsymbol{\mathrm{r}}_{M-1, n}$, $\nabla_p \boldsymbol{\mathrm{r}}^{\sigma} = \nabla_p \boldsymbol{\mathrm{r}}_{M-2, n} + \nabla_p \boldsymbol{\mathrm{r}}_{M-1, n}$, $\epsilon^{\sigma} = \epsilon_{M-2,n-1} + \epsilon_{M-1,n-1}$, $\frac{\partial\epsilon^{\sigma}}{\partial p} = \frac{\partial\epsilon_{M-2,n-1}}{\partial p} + \frac{\partial\epsilon_{M-1,n-1}}{\partial p}$
%where $\nabla_p \boldsymbol{\mathrm{r}}_{M-1,0}$ can be simplified into
%\begin{equation}
%\begin{aligned}
%\nabla_p \boldsymbol{\mathrm{r}}_{M-1,n} &= (\boldsymbol{I} - W_{M-1}^{-1} p \boldsymbol{\mathrm{A}}_{r,r,m,n})^{-1} \bigg[ W_{M-1}^{-1} \boldsymbol{\mathrm{A}}_{r,r,m,n} ( \nabla_p \boldsymbol{\mathrm{r}}_{M-2, n} + \nabla_p \boldsymbol{\mathrm{r}}_{M-1, n} + p \nabla_p \boldsymbol{\mathrm{r}}_{M-2, n} ) + W_n^{-1} \boldsymbol{\mathrm{A}}_{d,r,m,n}
%\\
%&( \nabla_p \boldsymbol{\mathrm{d}}_{M-1, n-1} + p \nabla_p \boldsymbol{\mathrm{d}}_{M-1, n-1}) + W_{M-1}^{-1} W_n^{-1} \Big( \frac{\partial \epsilon_{M-2,n-1}}{\partial p} + \frac{\partial \epsilon_{M-1,n-1}}{\partial p} \Big) \boldsymbol{\mathrm{u}}_{n,n} \bigg]
%\end{aligned}
%\end{equation}

\textbf{Case 6}: For $m = n = M-1$, we have
%\begin{equation}
%\begin{aligned}
%\label{eq:dste_updating6}
%\nabla_p \boldsymbol{\mathrm{r}}_{M-1,M-1} &= W_{M-1}^{-1} (\boldsymbol{\mathrm{A}}_{r,r,M-1,M-1} + \boldsymbol{\mathrm{A}}_{d,r,M-1,M-1}) (\nabla_p \boldsymbol{\mathrm{d}}_{M-1, M-2} + \nabla_p \boldsymbol{\mathrm{d}}_{M-1, M-1} + p \nabla_p \boldsymbol{\mathrm{d}}_{M-1, M-2} +
%\\
%&p \nabla_p \boldsymbol{\mathrm{d}}_{M-1, M-1}) + W_{M-1}^{-2} \bigg(\frac{\partial \epsilon_{M-2,M-2}}{\partial p} + \frac{\partial \epsilon_{M-2,M-1}}{\partial p} + \frac{\partial \epsilon_{M-1,M-2}}{\partial p} + \frac{\partial \epsilon_{M-1,M-1}}{\partial p}\bigg) \boldsymbol{\mathrm{u}}_{M-1,M-1}
%\\
%\nabla_p \boldsymbol{\mathrm{d}}_{M-1,M-1} &= \nabla_p \boldsymbol{\mathrm{r}}_{M-1,M-1}
%\\
%\frac{\partial \epsilon_{M-1,M-1}}{\partial p} &= 2 W_{M-1}^{-1} \boldsymbol{1}^T_{W_{M-1}-1} (\nabla_p \boldsymbol{\mathrm{d}}_{M-1, M-2} + \nabla_p \boldsymbol{\mathrm{d}}_{M-1, M-1} + p \nabla_p \boldsymbol{\mathrm{d}}_{M-1, M-2} + p \nabla_p \boldsymbol{\mathrm{d}}_{M-1, M-1})
%\\
%&+ W_{M-1}^{-1} \bigg(\frac{\partial \epsilon_{M-2,M-2}}{\partial p} + \frac{\partial \epsilon_{M-2,M-1}}{\partial p} + \frac{\partial \epsilon_{M-1,M-2}}{\partial p} + \frac{\partial \epsilon_{M-1,M-1}}{\partial p}\bigg)
%\end{aligned}
%\end{equation}
%where we have that
\begin{equation}
\begin{aligned}
\label{eq:dste_updating6}
\frac{\partial \epsilon_{M-1,M-1}}{\partial p}  &= \Big[ 1 - 2 W_{M-1}^{-3} p \boldsymbol{1}^T_{W_{M-1}-1} \big( \boldsymbol{I} - W_{M-1}^{-1} p \boldsymbol{A}^{\sigma} \big)^{-1} \boldsymbol{u}_{M-1,M-1} - W_{M-1}^{-1} \Big]^{-1}
\bigg\{ 2 W_{M-1}^{-1} \boldsymbol{1}^T_{W_{M-1}-1} (\nabla_p \boldsymbol{\mathrm{d}}_{M-1, M-2} +
\\
&\nabla_p \boldsymbol{\mathrm{d}}_{M-1, M-1} + p \nabla_p \boldsymbol{\mathrm{d}}_{M-1, M-2}) + W_{M-1}^{-1} \frac{\partial \epsilon^{\sigma}}{\partial p} + 2 W_{M-1}^{-1} p \boldsymbol{1}^T_{W_{M-1}-1} \big( \boldsymbol{I} - W_{M-1}^{-1} p \boldsymbol{A}^{\sigma} \big)^{-1} \Big[ W_{M-1}^{-1} \boldsymbol{A}^{\sigma}
\\
&(\nabla_p \boldsymbol{\mathrm{d}}_{M-1, M-2} + \nabla_p \boldsymbol{\mathrm{d}}_{M-1, M-1}
+ p \nabla_p \boldsymbol{\mathrm{d}}_{M-1, M-2}) + W_{M-1}^{-2} \frac{\partial \epsilon^{\sigma}}{\partial p}  \boldsymbol{u}_{M-1,M-1} \Big] \bigg\},
\\
\nabla_p \boldsymbol{\mathrm{r}}_{M-1,M-1} &= \nabla_p \boldsymbol{\mathrm{d}}_{M-1,M-1}
\\
&= \big( \boldsymbol{I} - W_{M-1}^{-1} p \boldsymbol{A}^{\sigma} \big)^{-1} \bigg[ W_{M-1}^{-1} p \boldsymbol{A}^{\sigma} ( \nabla_p \boldsymbol{\mathrm{d}}_{M-1, M-2} + \nabla_p \boldsymbol{\mathrm{d}}_{M-1, M-1} + p \nabla_p \boldsymbol{\mathrm{d}}_{M-1, M-2})
\\
&+ W_{M-1}^{-2} \Big( \frac{\partial \epsilon^{\sigma}}{\partial p} + \frac{\partial \epsilon_{M-1,M-1}}{\partial p} \Big) \boldsymbol{u}_{M-1,M-1} \bigg],
\end{aligned}
\end{equation}
where $\boldsymbol{A}^{\sigma} = \boldsymbol{A}_{r,r,M-1,M-1} + \boldsymbol{A}_{r,d,M-1,M-1} = \boldsymbol{A}_{d,d,M-1,M-1} + \boldsymbol{A}_{d,r,M-1,M-1}$; and
\begin{equation}
\begin{aligned}
\frac{\partial \epsilon^{\sigma}}{\partial p} = \frac{\partial \epsilon_{M-2,M-2}}{\partial p} + \frac{\partial \epsilon_{M-2,M-1}}{\partial p} + \frac{\partial \epsilon_{M-1,M-2}}{\partial p}.
\end{aligned}
\end{equation}

\textbf{Case 7}: For $0 \leq m < n \leq M-1$, we have that $\nabla_p \boldsymbol{\mathrm{r}}_{m,n} = \nabla_p \boldsymbol{\mathrm{d}}_{n,m}, \nabla_p \boldsymbol{\mathrm{d}}_{m,n} = \nabla_p \boldsymbol{\mathrm{r}}_{n,m}, \frac{\partial \epsilon_{m,n}}{\partial p} = \frac{\partial \epsilon_{n,m}}{\partial p}$.

\end{Lemma}

%\emph{Proof:} Please refer to Appendix D.

According to Lemma 1, we can utilize $\epsilon_{0,0}$ and $\frac{\partial \epsilon_{0,0}}{\partial p}$ to obtain $\nabla_p \boldsymbol{\mathrm{r}}_{m,n}, \nabla_p \boldsymbol{\mathrm{d}}_{m,n}$ and $\frac{\partial \epsilon_{m,n}}{\partial p}$ for $0 \leq m,n \leq M-1$.
Furthermore, the normalization condition of $\nabla_p \boldsymbol{\mathrm{r}}_{m,n}, \nabla_p \boldsymbol{\mathrm{d}}_{m,n}$ and $\frac{\partial \epsilon_{m,n}}{\partial p}$ is given by the following equation,
\begin{equation}
\begin{aligned}
\label{eq:dsigma_normalizing_condition}
\sum_{m=0}^{M-1} \sum_{n=0}^{M-1} \frac{\partial \mathbb{P}(m,n)}{\partial p} = \frac{\partial}{\partial p} \sum_{m=0}^{M-1} \sum_{n=0}^{M-1} \mathbb{P}(m,n) = 0.
\end{aligned}
\end{equation}
Therefore, we propose the following Lemma 2 to calculate $\frac{\partial \mathbb{P}(m,n)}{\partial p}$ using $\nabla_p \boldsymbol{\mathrm{r}}_{m,n}, \nabla_p \boldsymbol{\mathrm{d}}_{m,n}$ and $\frac{\partial \epsilon_{m,n}}{\partial p}$ for $0 \leq m,n \leq M-1$.

\begin{Lemma}
Based on the $7$ cases in Theorem 2, the partial derivative $\frac{\partial \mathbb{P}(m,n)}{\partial p}$ for $0 \leq m,n \leq M-1$ can be computed as follows,

\textbf{Case 1}: For $m = n = 0$, we have
\begin{equation}
\begin{aligned}
\label{eq:dsigma_r_d1}
\frac{\partial \mathbb{P}(0,0)}{\partial p} = \frac{1}{6} (2 W_0 + 1) (W_0 + 1) \frac{\partial \epsilon_{0,0}}{\partial p}.
\end{aligned}
\end{equation}

\textbf{Case 2}: For $0 < m < M-1$, and $n = 0$, we have
\begin{equation}
\begin{aligned}
\label{eq:dsigma_r_d2}
\frac{\partial \mathbb{P}(m,0)}{\partial p} = W_m^{-1} \sum^{W_0-1}_{i=1} \Big( k \frac{\partial}{\partial p} r_{m-1,0,i} + p \frac{\partial r_{m-1,0,i}}{\partial p} \Big) \bigg[ -\frac{1}{2} i^2 + \Big(W_m + \frac{1}{2}\Big)i \bigg].
\end{aligned}
\end{equation}

\textbf{Case 3}: For $m = M-1$ and $n = 0$, we have
\begin{equation}
\begin{aligned}
\label{eq:dsigma_r_d3}
\frac{\partial \mathbb{P}(M-1,0)}{\partial p} = W_{M-1}^{-1} \sum^{W_0-1}_{i=1} \Big(k \frac{\partial}{\partial p} r_{M-2,0,i} \!+\! k \frac{\partial}{\partial p} r_{M-1,0,i} \!+\! p \frac{\partial r_{M-2,0,i}}{\partial p} \!+\! p \frac{\partial r_{M-1,0,i}}{\partial p}\Big) \bigg[ -\frac{1}{2} i^2 \!+\! \Big(W_{M-1} \!+\! \frac{1}{2}\Big)i \bigg].
\end{aligned}
\end{equation}

\textbf{Case 4}: For $0 < n \leq m < M-1$, we have
\begin{equation}
\begin{aligned}
\label{eq:dsigma_r_d4}
\frac{\partial \mathbb{P}(m,n)}{\partial p} &= W_m^{-1} \sum^{W_n-1}_{i=1} \Big( k \frac{\partial}{\partial p} r_{m-1,n,i} \!+\! p \frac{\partial r_{m-1,n,i}}{\partial p} \Big) \bigg[ -\frac{1}{2} i^2 \!+\! \Big(W_m + \frac{1}{2}\Big)i \bigg] \!+\! W_n^{-1} \sum^{W_n-1}_{i=1} \Big( k \frac{\partial}{\partial p} d_{m,n-1,i} \!+\! p \frac{\partial d_{m,n-1,i}}{\partial p} \Big)
\\
&\bigg[ -\frac{1}{2} i^2 + \Big(W_n + \frac{1}{2}\Big)i \bigg] + \Big( -\frac{1}{2} W_n + \frac{1}{2} \Big) \sum^{W_m-1}_{i=W_n} \Big( k \frac{\partial}{\partial p} d_{m,n-1,i} + p \frac{\partial d_{m,n-1,i}}{\partial p} \Big) + W_m^{-1} W_n^{-1} \frac{\partial \epsilon_{m-1,n-1}}{\partial p}
\\
&\Big( -\frac{1}{6}W_n^3 + \frac{1}{2} W_m W_n^2 + \frac{1}{2} W_m W_n + \frac{1}{6} W_n \Big).
\end{aligned}
\end{equation}

\textbf{Case 5}: For $m = M-1$ we have $0 < n < M-1$, we have
\begin{equation}
\begin{aligned}
\label{eq:dsigma_r_d5}
&\frac{\partial \mathbb{P}(M-1,n)}{\partial p} = W_{M-1}^{-1}  \sum^{W_n-1}_{i=1} \Big( k \frac{\partial}{\partial p} r_{M-2,n,i} \!+\! k \frac{\partial}{\partial p} r_{M-1,n,i} \!+\! p \frac{\partial r_{M-2,n,i}}{\partial p} \!+\! p \frac{\partial r_{M-1,n,i}}{\partial p} \Big) \bigg[ -\! \frac{1}{2} i^2 \!+\! \Big(W_{M-1} \!+\! \frac{1}{2}\Big)i \bigg]
\\
&+ W_n^{-1} \sum^{W_n-1}_{i=1} \Big( k \frac{\partial}{\partial p} d_{M-1,n-1,i} + p \frac{\partial d_{M-1,n-1,i}}{\partial p} \Big) \bigg[ -\frac{1}{2} i^2 + \Big(W_n + \frac{1}{2}\Big)i \bigg] + \Big( -\frac{1}{2} W_n + \frac{1}{2} \Big) \sum^{W_{M-1}-1}_{i=W_n} \Big( k \frac{\partial}{\partial p} d_{M-1,n-1,i}
\\
&+ p \frac{\partial d_{M-1,n-1,i}}{\partial p} \Big) + W_{M-1}^{-1} W_n^{-1} \Big( \frac{\partial \epsilon_{M-2,n-1}}{\partial p} + \frac{\partial \epsilon_{M-1,n-1}}{\partial p} \Big) \Big( -\frac{1}{6}W_n^3 + \frac{1}{2} W_{M-1} W_n^2 + \frac{1}{2} W_{M-1} W_n + \frac{1}{6} W_n \Big).
\end{aligned}
\end{equation}

\textbf{Case 6}: For $m = n = M-1$, we have
\begin{equation}
\begin{aligned}
\label{eq:dsigma_r_d6}
\frac{\partial \mathbb{P}(M-1,M-1)}{\partial p} &= W_{M-1}^{-1} \sum^{W-1}_{i=1} \Big( k \frac{\partial}{\partial p} d^{\sigma}_i \!+\! p \frac{\partial d^{\sigma}_i}{\partial p} \Big) \bigg[ -\frac{1}{2} i^2 \!+\! \Big(W_{M-1} \!+\! \frac{1}{2}\Big)i \bigg] \!+\! \frac{\partial \epsilon^{\sigma}}{\partial p} \Big(\frac{1}{3}W_{M-1} \!+\! \frac{1}{2} \!+\! \frac{1}{6}W_{M-1}^{-1} \Big),
\nonumber
\end{aligned}
\end{equation}
where
\begin{equation}
\begin{aligned}
%\label{eq:dsigma_r_d6}
\frac{\partial d^{\sigma}_i}{\partial p} &= \frac{\partial r_{m-1,n,i}}{\partial p} + \frac{\partial r_{m,n,i}}{\partial p} + \frac{\partial d_{m,n-1,i}}{\partial p} + \frac{\partial d_{m,n,i}}{\partial p}
\\
\frac{\partial \epsilon^{\sigma}}{\partial p} &= \frac{\partial \epsilon_{M-2,M-2}}{\partial p} + \frac{\partial \epsilon_{M-2,M-1}}{\partial p} + \frac{\partial \epsilon_{M-1,M-2}}{\partial p} + \frac{\partial \epsilon_{M-1,M-1}}{\partial p}.
\end{aligned}
\end{equation}

\textbf{Case 7}: For $0 \leq n < m \leq M-1$, we have that $\frac{\partial \mathbb{P}(m,n)}{\partial p} = \frac{\partial \mathbb{P}(n,m)}{\partial p}$.

\end{Lemma}
%\emph{Proof:} Please refer to Appendix ???.
%

Based on Lemmas 1 and 2, we can use $\epsilon_{0,0}$ and $\frac{\partial \epsilon_{0,0}}{\partial p}$ to characterize $\frac{\partial \mathbb{P}(m,n)}{\partial p}$, and utilize the normalization condition give by Equation (\ref{eq:dsigma_normalizing_condition}) to obtain $\frac{\partial \epsilon_{0,0}}{\partial p}$.
Furthermore, we can calculate $\nabla_p \boldsymbol{\mathrm{r}}_{m,n}, \nabla_p \boldsymbol{\mathrm{d}}_{m,n}$ and $\frac{\partial \epsilon_{m,n}}{\partial p}$ for $0 \leq m,n \leq M-1$ by $\frac{\partial \epsilon_{0,0}}{\partial p}$ based on Lemma 1, and achieve $\frac{\partial \eta}{\partial p}$ according to Equation (\ref{eq:eta_derivate_p}).
Therefore, we can further calculate $p$ according to Equation (\ref{eq:p_updating}), where the key variables are shown in Table \ref{tb:vertor_variables}.
% In summary, based on Theorems 1, 2, Lemmas 1,2 and Newton method, we give the necessary variables to calculate numerical $\hat{p}$ in Table \ref{tb:vertor_variables}, and propose the Algorithm \ref{ag:calculate_p} to calculate $\hat{p}$ numerically.

\begin{table}[tbp]
	\footnotesize
	\caption{The expression of vector form of state probabilities.}
	\label{tb:vertor_variables}
	\centering
	\begin{tabular}{|c|c|c|c|c|c|c|}  % {lccc} ��ʾ����Ԫ�ض��뷽ʽ��left-l,right-r,center-c
		\hline
		& State transition & State probability & $\boldsymbol{\mathrm{r}}_{m,n}, \boldsymbol{\mathrm{d}}_{m,n}, \epsilon_{m,n}$ & $\mathbb{P}(m,n)$ & $\nabla_p \boldsymbol{\mathrm{r}}_{m,n}, \nabla_p \boldsymbol{\mathrm{d}}_{m,n}, \frac{\partial \epsilon_{m,n}}{\partial p}  $ & $\frac{\partial \mathbb{P}(m,n)}{\partial p}$ \\
		\hline
		$m=n=0$ & Fig. \ref{fig:m=0,n=0} & (\ref{eq:ste_Pr1}) & (\ref{eq:ste_updating1}) & (\ref{eq:sigma_r_d1}) & (\ref{eq:dste_updating1}) & (\ref{eq:dsigma_r_d1}) \\
		\hline
		$m=M-1,n=0$ & Fig. \ref{fig:n=0} & (\ref{eq:ste_Pr2}) & (\ref{eq:ste_updating2}) & (\ref{eq:sigma_r_d2}) & (\ref{eq:dste_updating2}) & (\ref{eq:dsigma_r_d2}) \\
		\hline
		$0<m<M-1,n=0$ & Fig. \ref{fig:m=M-1,n=0} & (\ref{eq:ste_Pr3}) & (\ref{eq:ste_updating3}) & (\ref{eq:sigma_r_d3}) & (\ref{eq:dste_updating3}) & (\ref{eq:dsigma_r_d3}) \\
		\hline
		$0<n\leq m<M-1$ & Fig. \ref{fig:m,n} & (\ref{eq:ste_Pr4}) & (\ref{eq:ste_updating4}) & (\ref{eq:sigma_r_d4}) & (\ref{eq:dste_updating4}) & (\ref{eq:dsigma_r_d4}) \\
		\hline
		$m=M-1,0<n<M-1$ & Fig. \ref{fig:n=M-1} & (\ref{eq:ste_Pr5}) & (\ref{eq:ste_updating5}) & (\ref{eq:sigma_r_d5}) & (\ref{eq:dste_updating5}) & (\ref{eq:dsigma_r_d5}) \\
		\hline
		$m=n=M-1$ & Fig. \ref{fig:m=M-1,n=M-1} & (\ref{eq:ste_Pr6}) & (\ref{eq:ste_updating6}) & (\ref{eq:sigma_r_d6}) & (\ref{eq:dste_updating6}) & (\ref{eq:dsigma_r_d6}) \\
		\hline
		$0<m<n<M-1$ & $--$ & (\ref{eq:ste_Pr7}) & $\boldsymbol{\mathrm{r}}_{n,m}, \boldsymbol{\mathrm{d}}_{n,m}, \epsilon_{n,m}$ & $\mathbb{P}(n,m)$ & $\nabla_p \boldsymbol{\mathrm{r}}_{n,m}, \nabla_p \boldsymbol{\mathrm{d}}_{n,m}, \frac{\partial \epsilon_{n,m}}{\partial p}  $ & $\frac{\partial \mathbb{P}(n,m)}{\partial p}$ \\
		\hline
	\end{tabular}
\end{table}

%\begin{algorithm}[t]
%\label{ag:calculate_p}
%\caption{Calculate $\hat{p}$}% �㷨����
%\LinesNumbered %Ҫ����ʾ�к�
%\KwIn{parameters $W_0, N, V$}% ��������
%\KwOut{$\hat{p}, \eta, \frac{\partial \eta}{\partial p}$}% ����
%$v \gets 0$, $\hat{p}^{(0)} \gets 0$ \;
%\While{$v < V$}{
%    Calculate $\boldsymbol{\mathrm{r}}_{m,n}$, $\boldsymbol{\mathrm{d}}_{m,n}$ and $\epsilon_{m,n}$ for $0 \leq m,n \leq M-1$ using Equations (\ref{eq:ste_updating1})$-$(\ref{eq:sigma_r_d6})\;
%    Calculate $\nabla_p \boldsymbol{\mathrm{r}}_{m,n}$, $\nabla_p \boldsymbol{\mathrm{d}}_{m,n}$ and $\frac{\partial \epsilon_{m,n}}{\partial p}$ for $0 \leq m,n \leq M-1$ using Equations (\ref{eq:dste_updating1})$-$(\ref{eq:dsigma_r_d6})\;
%    Calculate $\eta|_{p=p^{(v)}}$ and $\frac{\partial \eta}{\partial p}|_{p=p^{(v)}}$ using Equations (\ref{eq:eta2}) and (\ref{eq:eta_derivate_p})\;
%	Calculate $\hat{p}^{(v+1)}$ using Equation (\ref{eq:p_updating})\;
%    $v \gets v + 1$\;
%}
%\textbf{return} $p^{(V)}, \eta|_{p=p^{(V)}}, \frac{\partial \eta}{\partial p}|_{p=p^{(V)}}$
%\end{algorithm}

\subsection{Numerical Solution of Throughput and Transmission Delay}

According to Section IV in \cite{bianchi2000performance}, the expectation of successful transmissions is given by $2 L_p N \eta(1-\eta)^{N-1}$; and that of total transmission equals $T_s N\eta(1-\eta)^{N-1} + T_c\big[1-N\eta(1-\eta)^{N-1}-(1-\eta)^N\big] + \tau(1-\eta)^N$,
where $L_p$ denotes the number of transmitted symbols within one data frame transmission; $T_s$ and $T_c$ denote the average channel busy due to a successful transmission and a collision, respectively, given by
\begin{equation}
\begin{aligned}
T_s &= \text{RTS-S + SIFS + PTA + SIFS + SAK + SIFS + CTS-S + SIFS }
\\
&+ \text{PHY-H + MAC-H + $L_p$ + SIFS + ACK + DIFS};
\\
T_c &= \text{PTA + DIFS};
\end{aligned}
\end{equation}
and PHY-H as well as MAC-H denote the header of physical and MAC layers, respectively.

Generally, let $C$ denote the  throughput given by the following equation,
\begin{equation}
\begin{aligned}
\label{eq:throughput_expression}
C = \frac{2 L_p N \eta(1-\eta)^{N-1}}{T_s N\eta(1-\eta)^{N-1} + T_c\big[1-N\eta(1-\eta)^{N-1}-(1-\eta)^N\big] + \tau(1-\eta)^N} = \frac{2 L_p}{T_s + \tau L_o^{-1} - T_c},
\end{aligned}
\end{equation}
where $\tau$ denotes the duration of once backoff; $L_p, T_s, T_c$ as well as $\tau$ have to be characterized by same unit; and $L_o$ is given by
\begin{equation}
\begin{aligned}
\label{eq:L_o}
L_{o} = \frac{N \eta (1-\eta)^{N-1}}{T_c \tau^{-1} - (1-\eta)^N(T_c \tau^{-1} - 1)}.
\end{aligned}
\end{equation}

Finally, the average transmission delay can be estimated by $D = \frac{L_p}{C}$.

%light velocity $c$, and the throughput in Equation (\ref{eq:throughput_expression}), the average transmission delay $C$ can be estimated by
%\begin{equation}
%\begin{aligned}
%\label{eq:trans_delay}
%D =  + \frac{L_p}{C},
%\end{aligned}
%\end{equation}
%%where $\delta = \frac{d}{c}$ denotes the transmission period of optical signal.
%It is implied that the minimization of transmission delay is equivalent to the maximization of throughput, hence, we only give the derivation of maximizing throughput in the following sections.

\section{Optimization for Initial Contention Window and indicator matrix}

\subsection{Optimization for Initial Contention Window}

We propose to optimize initial contention window $W_0^*$ to maximize the system throughput $C$ given the number of TCPairs $N$ in this subsection.
According to Equations (\ref{eq:throughput_expression}) and (\ref{eq:L_o}), $C$ depends on $\eta$ and thus depends on $W_0$, where $W_0 \in 2^{\mathbb{N}^{+}}$, and $2^{\mathbb{N}^{+}} \triangleq \{ 2^{1}, 2^{2}, \cdots \}$.
%Generally, the maximum throughput $C^{*}$ is available when $\eta = \arg\max C \triangleq \eta^{*} $, where $W_0 = \arg \{ \eta = \eta^* \} \triangleq W_0^*$.
In order to optimize $W_0$ in a tractable manner, we slack $W_0$ from $2^{\mathbb{N}^{+}}$ to $\mathbb{R}$, and obtain the optimal $\widetilde{W}_0^* = \arg \max\limits_{\widetilde{W}_0 \in \mathbb{R}} C$.
Afterwards, based on $W_{0,l}, W_{0,r} \in 2^{\mathbb{N}^{+}}$, the most two adjacent values to the optimal $\widetilde{W}_0^*$, we have that $W_0^* = \arg \!\!\!\!\!\!\! \max\limits_{W_0 \in \{ W_{0,l}, W_{0,r} \}} \!\!\!\!\!\!\! C$.
Generally, $\widetilde{W}_0^*$ can be approximatively estimated from Theorem 3.

\begin{Theorem}
The optimal slacked initial contention window $\widetilde{W}_0^*$ can be approximated by
\begin{equation}
\begin{aligned}
\label{eq:optimal_W_0}
\widetilde{W}^*_0 \approx \frac{3}{2 \sqrt{2}} N \gamma - \frac{3}{4} + \sqrt{\frac{9}{8} N^2 \gamma^2 - \frac{21 \sqrt{2}}{8} N \gamma + \frac{1}{16}}
\end{aligned}
\end{equation}
where $\gamma = \sqrt{T_c \tau^{-1}}$, and $T_c$ as well as $\tau$ are denoted in Section IV.B.
\end{Theorem}

\emph{Proof}: Please refer to Appendix E.

\subsection{Optimization for TCPair Number}

For the optimization of indicator matrix, we propose to optimize the TCPair number to maximize system throughput $C$ given initial contention window $W_0$.
Via slacking $N$ from $\mathbb{N}^{+}$ to $\mathbb{R}$, we can obtain the optimal $\widetilde{N}^*$ that satisfies $\widetilde{N}^* = \arg \max\limits_{\widetilde{N} \in \mathbb{R}} C$, and its two adjacent integral neighbours $N_{l}$ and $N_{r}$.
Then, we select the optimal $N^*$ by $N^* = \arg \max\limits_{N \in \{ N_{l}, N_{r} \}} C$.
Generally, the optimal $\widetilde{N}^*$ can be approximatively calculated according to Theorem 4.
%According to Equation (\ref{eq:throughput_expression}), the maximization of $C$ is equivalent to maximizing $L_o$, which means the optimal $\widetilde{N}^*$ satisfies $N^* = \arg\max C = \arg\max L_o = \arg \big\{ \frac{\partial L_o}{\partial N}=0 \big\}$.

\begin{Theorem}
The approximated slacked optimal number of TCPairs $\widetilde{N}^*$ is given by
\begin{equation}
\begin{aligned}
\label{eq:optimal_N}
\widetilde{N}^* \approx \bigg(\frac{T_c}{\tau} - 1\bigg)^{-1} \Bigg[ \sqrt{\frac{\eta^2}{4} + \eta + \frac{2 T_c}{\eta} - 1} - \bigg( 1 + \frac{\eta}{2} \bigg) \Bigg] \eta^{-1},
\end{aligned}
\end{equation}
where $\eta = (W_0 - 1) \big[ \frac{1}{3} W_0^2 + \frac{1}{2} W_0 + \frac{1}{6} \big]^{-1}$.
\end{Theorem}

\emph{Proof}: Please refer to Appendix F.

\subsection{Optimization for the Indicator Matrix}

% Due to the length limit of manuscript, please refer to [Arxiv site] for details.

\textcolor{red}{
Assuming that the system's physical-layer-connectivity is characterized in an adjacent matrix $\boldsymbol{S} \in \{0,1\}^{N_s \times N_s}$ ($\boldsymbol{S} = \boldsymbol{S}^T$), we optimize the indicator matrix $\boldsymbol{\Phi}$ subjected to the optimal number of TCPs $N^*$ in order to balance different stations's chances to transmit packages.
Hence, we minimize the variance of different stations' PTCouter numbers, i. e.,
\begin{equation}
\begin{aligned}
\label{eq:confined_optimization}
&\boldsymbol{\Phi}^* = \arg \!\!\!\!\! \min\limits_{\boldsymbol{\Phi}\in\{0,1\}^{N_s\times N_s}} \sum_{i=1}^{N_s} \Big( \sum_{j=1}^{N_s} \phi_{i,j} \Big)^2
% = \arg\min\limits_{\boldsymbol{\Phi}\in\{0,1\}^{N_s\times N_s}} \mathbb{D}(J_1, J_2, \cdots, J_{N_s}),
\\
&\text{s. t.} \sum^{N_s}_{i=1}\sum^{N_s}_{j=1}\phi_{i,j} = N^*; \boldsymbol{\Phi} = \boldsymbol{\Phi}^T, \boldsymbol{\Phi} \circ \bS = \boldsymbol{\Phi},
\end{aligned}
\end{equation}
where $\boldsymbol{\Phi} \circ \bS$ denotes the element-wise product of matrices $\boldsymbol{\Phi}$ and $\bS$.
For the simplicity, we let $\mathcal{R}_i(\bS) = \sum^{N_s}_{j=1} s_{i,j}$ and $Q(\bS) = \sum^{N_s}_{i=1} \big[ \mathcal{R}_i(\bS) \big]^2$,
the objective function is thus to be $\boldsymbol{\Phi}^* =  \arg \!\!\!\!\!\!\! \min\limits_{\boldsymbol{\Phi} \in \{0,1\}^{N_s\times N_s}} \!\!\!\!\!\!\! Q(\boldsymbol{\Phi}).$}
% where $J_i$ denotes the number of PTCounters of station $i$ (also the summation of column $i$ in matrix $\boldsymbol{\Phi}$); $\mathbb{D}[\bullet]$ denotes the variance function;
%\begin{equation}
%\begin{aligned}
%\boldsymbol{\Phi} =  \arg\min\limits_{\boldsymbol{\Phi} \in \{0,1\}^{N_s\times N_s}} Q(\boldsymbol{\Phi}).
%\end{aligned}
%\end{equation}

Let $\mathcal{S}_0 = \{ \bS \}$ and $\mathcal{B}^v (\mathcal{S}_0)$ denote the set given by
\begin{equation}
\begin{aligned}
\label{eq:B_v}
% \mathcal{B}^v (\mathcal{S}_0)
\mathcal{S}_v = \Big\{ \boldsymbol{\Phi} \mid \boldsymbol{\Phi} \in \{ 0,1\}^{N_s\times N_s}, \boldsymbol{\Phi} = \boldsymbol{\Phi}^T, \boldsymbol{\Phi} \circ \bS = \boldsymbol{\Phi}, \mathbb{D}(\boldsymbol{\Phi}, \bS) = 2v, \bS \in \mathcal{S}_0 \Big\},
\end{aligned}
\end{equation}
where $\mathbb{D}(\boldsymbol{\Phi}, \bS) = \sum^{N_s}_{i=1}\sum^{N_s}_{j=1} (\phi_{i,j} - s_{i,j})^2$ denotes the Hamming distance between matrices $\boldsymbol{\Phi}$ and $\bS$.
Therefore, we can convert the constrained optimization into $\boldsymbol{\Phi}^* = \arg \min\limits_{\boldsymbol{\Phi} \in \mathcal{S}_v} Q(\boldsymbol{\Phi})$, and resort a greedy method given by the following equation to approximately minimize $Q(\boldsymbol{\Phi})$,
\begin{equation}
\begin{aligned}
\label{eq:dp1}
\min\limits_{\boldsymbol{\Phi} \in \mathcal{S}_v} Q(\boldsymbol{\Phi})
\approx \min\limits_{\boldsymbol{\Phi} \in \mathcal{S}_{v-1}} \Big\{ Q(\boldsymbol{\Phi}) - \max\limits_{1\leq i,j \leq N_s} s_{i,j} \big\{ \mathcal{R}_i(\boldsymbol{\Phi}) + \mathcal{R}_j(\boldsymbol{\Phi}) \big\} \Big\} + 2.
%\\
%&:= \min\limits_{\boldsymbol{\Phi} \in \mathcal{B}^{v-1} (\mathcal{S}_0)}  Q(\boldsymbol{\Phi}) - \max\limits_{1\leq i,j \leq N_s} s_{i,j} \big\{ \mathcal{R}_i(\boldsymbol{\Phi}^*) + \mathcal{R}_j(\boldsymbol{\Phi}^*) \big\} + 2,
\end{aligned}
\end{equation}
%where $\boldsymbol{\Phi}^* = \arg \min\limits_{\boldsymbol{\Phi} \in \mathcal{B}^{v-1} (\mathcal{S}_0)} Q(\boldsymbol{\Phi})$.
%\emph{Proof}: Please refer to Appendix F.
Furthermore,
to reduce the solution space, we let $\mathcal{S}_v$ only include step-optimal $\boldsymbol{\Phi}$, i.e. $\mathcal{S}_v = \{ \boldsymbol{\Phi} \mid \boldsymbol{\Phi} = \arg \min Q(\boldsymbol{\Phi}) \}$, and
%\begin{equation}
%\begin{aligned}
%\label{eq:s_v}
%\mathcal{S}_v = \big\{ \boldsymbol{\Phi} \mid Q(\boldsymbol{\Phi}) = \min_{\boldsymbol{\Phi}\in\mathcal{B}^v (\mathcal{S}_0)} Q(\boldsymbol{\Phi}) \big\},
%\end{aligned}
%\end{equation}
transform Equation (\ref{eq:dp1}) into the following recursion equation,
\begin{equation}
\begin{aligned}
\label{eq:dp2}
\min\limits_{\boldsymbol{\Phi} \in \mathcal{S}_v} Q(\boldsymbol{\Phi}) \approx \min\limits_{\boldsymbol{\Phi} \in \mathcal{S}_{v-1}} Q(\boldsymbol{\Phi}) - \max\limits_{\boldsymbol{\Phi} \in \mathcal{S}_{v-1}} \max\limits_{1\leq i,j \leq N_s} s_{i,j} \big\{ \mathcal{R}_i(\boldsymbol{\Phi}) + \mathcal{R}_j(\boldsymbol{\Phi}) \big\} + 2.
\end{aligned}
\end{equation}
Consequently, we have a recursive expression on $\mathcal{S}_{v+1}$ given as follow,
\begin{equation}
\begin{aligned}
\label{eq:S_v+1}
\mathcal{S}_{v} = \Big\{ \boldsymbol{\Phi} - \boldsymbol{0}_{i,j} - \boldsymbol{0}_{j,i} \mid \boldsymbol{\Phi}, i, j = \arg \max\limits_{\boldsymbol{\Phi}\in \mathcal{S}_{v-1}} \max\limits_{1 \leq i,j \leq N_s} \mathcal{R}_i(\boldsymbol{\Phi}) + \mathcal{R}_j(\boldsymbol{\Phi}) \Big\},
%\in \mathcal{B} (\mathcal{S}_v), \max\limits_{1 \leq i,j \leq N_s} \mathcal{R}_i(\boldsymbol{\Phi}) + \mathcal{R}_j(\boldsymbol{\Phi})
% g_{\hat{i}, \hat{j}}(\hat{\boldsymbol{\Phi}}) = g^*_v, (\hat{\boldsymbol{\Phi}}, \hat{i}, \hat{j})
% = \mathcal{B}^{-1} \big[ \mathcal{B} (\mathcal{S}_v) \big] \Big\},
\end{aligned}
\end{equation}
where $\boldsymbol{0}_{i,j}$ denotes a full-zero matrix with only element $(i,j)$ equals $1$.
%where $g_{i,j} (\boldsymbol{\Phi}) = \mathcal{R}_i(\boldsymbol{\Phi}) + \mathcal{R}_j(\boldsymbol{\Phi})$; $g^*_v = \max\limits_{\boldsymbol{\Phi} \in \mathcal{S}_v} \max\limits_{1\leq i,j \leq N_s} s_{i,j} \big[ \mathcal{R}_i(\boldsymbol{\Phi}) + \mathcal{R}_j(\boldsymbol{\Phi}) \big]$; and $\mathcal{B}^{-1} \big[ \mathcal{B} (\mathcal{S}_v) \big]$ is given as follows,
%\begin{equation}
%\begin{aligned}
%\label{eq:B-1}
%\mathcal{B}^{-1} \big[ \mathcal{B} (\mathcal{S}_v) \big] = \Big\{ (\hat{\boldsymbol{\Phi}}, \hat{i}, \hat{j}) | \hat{\boldsymbol{\Phi}} \in \mathcal{S}_v, \boldsymbol{\Phi} \in \mathcal{B} (\mathcal{S}_v), s_{\hat{i}, \hat{j}} = s_{\hat{j}, \hat{i}} = 0, \hat{s}_{\hat{i}, \hat{j}} = \hat{s}_{\hat{j}, \hat{i}} = 1 \Big\}.
%\end{aligned}
%\end{equation}

The iterative procedure starts from $v=0$ and terminates as $v = N - N^*$, and $\mathcal{S}_{N - N^*}$ exactly includes all possible optimal indicator matrices $\boldsymbol{\Phi}^*$ satisfying Equation (\ref{eq:confined_optimization}), which can balance all the stations' chances of data transmission.

\section{Numerical and Simulation Results}

We implement simulation and numerical results to evaluate the proposed DS-CSMA protocol.
The MAC-layer parameters are $M = 4$, $L_p = 8184$, MAC-H = 272, PHY-H = 128, RTS-S = 288, PTA = 240, SAK = 160, CTS-S = 160, ACK = 240, SIFS = 28, DIFS = 128 and $\tau = 50$, all in the unit of symbol duration which is assumed as $10$Mbps; all data package transmissions are based on superimposed transmission and can be referred to the channel model in \cite{wang2018characterization, Guanchu8845642}.

Figures \ref{fig:collistion_pro} and \ref{fig:throughput} plot the collision probability and overall throughput versus initial contention window length $W_0$, respectively, where the indicator matrix satisfies $\frac{1}{2} \sum_{i=1}^{N_s} \sum_{j=1}^{N_s} \phi_{i,j} = N$.
%\textcolor{red}{and the throughput of conventional CSMA/CA and multi-package reception (MPR) \cite{Babich2010} are shown for comparison.}
It is demonstrated that the proposed analytical model for DS-CSMA protocol is accurate enough since the numerical results approach the simulation one.
Furthermore, compared with CSMA/CA in IEEE 802.11/15 protocol, it is observed that our proposed DS-CSMA can almost double the throughput, where the employment of superimposed transmission plays a key role.

\begin{figure}
	\begin{minipage}[t]{0.49\textwidth}
	\centering
	\includegraphics[width=0.95\textwidth]{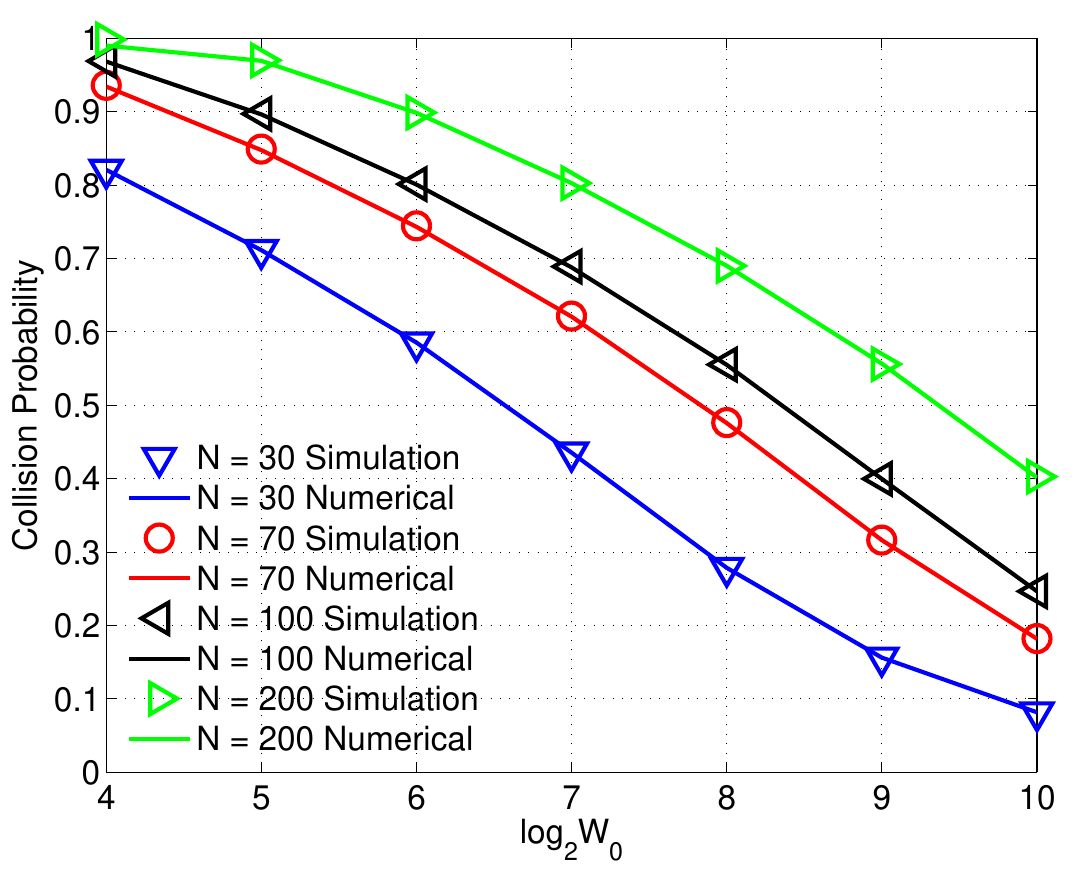}
    \caption{\label{fig:collistion_pro} The numerical and simulation result of collision probability.}
	\end{minipage}%
	\begin{minipage}[t]{0.49\textwidth}
	\centering
	\includegraphics[width=0.95\textwidth]{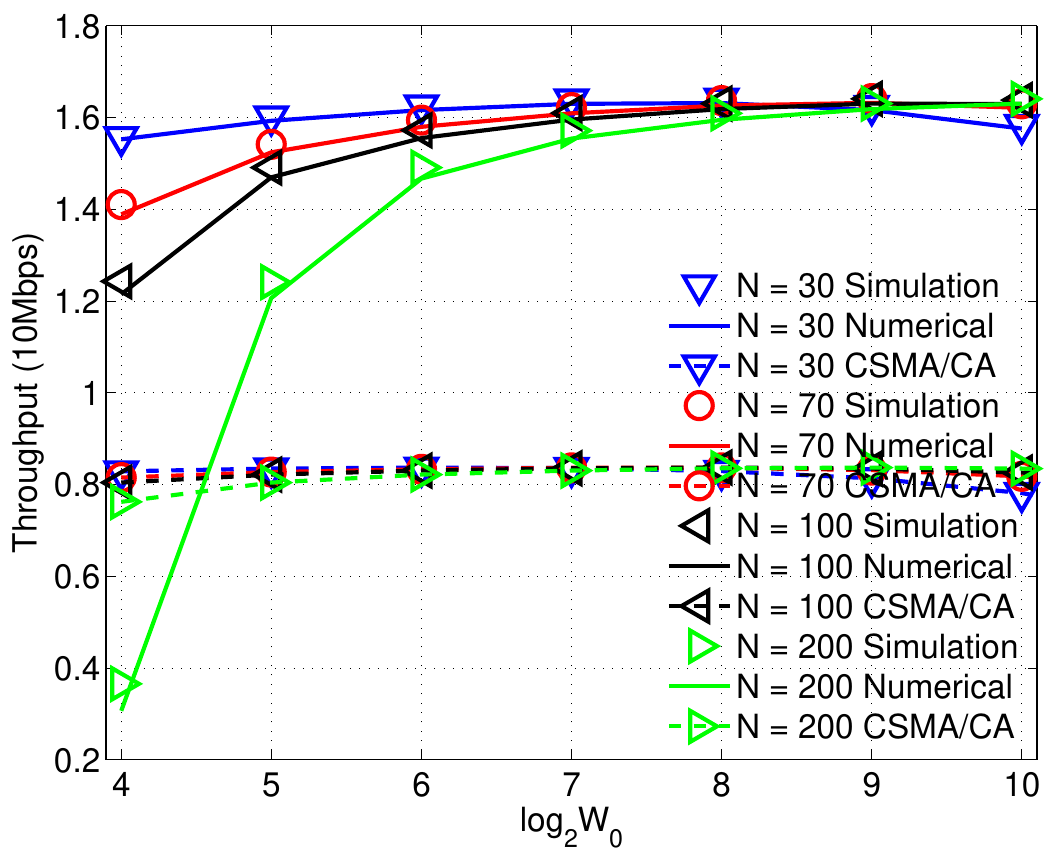}
	\caption{\label{fig:throughput} The numerical and simulation result of throughput.}
	\end{minipage}%
\end{figure}

Figures \ref{fig:W_opt} and \ref{fig:N_opt} illustrate the overall throughput versus the number of TCPairs $N$ and initial contention window length $W_0$, respectively.
It is observed that for fixed contention window, the optimal number of TCPairs are varied, and vice versa.
\textcolor{red}{
For small $N$, the throughput is lower since there are not enough TCPairs transmitting data frames; for too large $N$, too many TCPairs contend the channel to transmit data frames.
Hence, the throughput does not monotonically increases with $N$, but reaches the peak value for certain optimal $N^*$.
For small $W_0$, the throughput is lower due to large collision probability, as shown in Figure 6; for too large $W_0$, even though the transmission success rate is increased, all stations are wasting too much time on the backoff process.
Hence, the throughput does not monotonously increases with $W_0$, but reaches the peak value for certain optimal $W_0^*$ given by Table \ref{tb:optimal_parameters}.
The validity of Equations \ref{eq:optimal_W_0} and \ref{eq:optimal_N} can be proved by Figures \ref{fig:W_opt} and \ref{fig:N_opt}, where the maximum throughput can always be obtained.}

\textcolor{red}{
We give the comparison of throughput and average transmission delay with that of CSMA/CA with MPR \cite{Babich2010, Dua2008} in Figures \ref{fig:throughput_vs_MPR} and \ref{fig:delay_vs_MPR}.
It is observed that our proposed DS-CSMA protocol has higher throughput and lower transmission delay, which can be justified as follows.
In our proposed DS-CSMA protocol, the optimal number of TCPairs $N$ can be achieved by Equation (\ref{eq:optimal_N}) for different $W_0$; while in MPR or other existing protocol $N$ equals the number of stations which only depends on the network structure and cannot be optimized.}

\begin{figure}
	\begin{minipage}[t]{0.49\textwidth}
	\centering
	\includegraphics[width=0.95\textwidth]{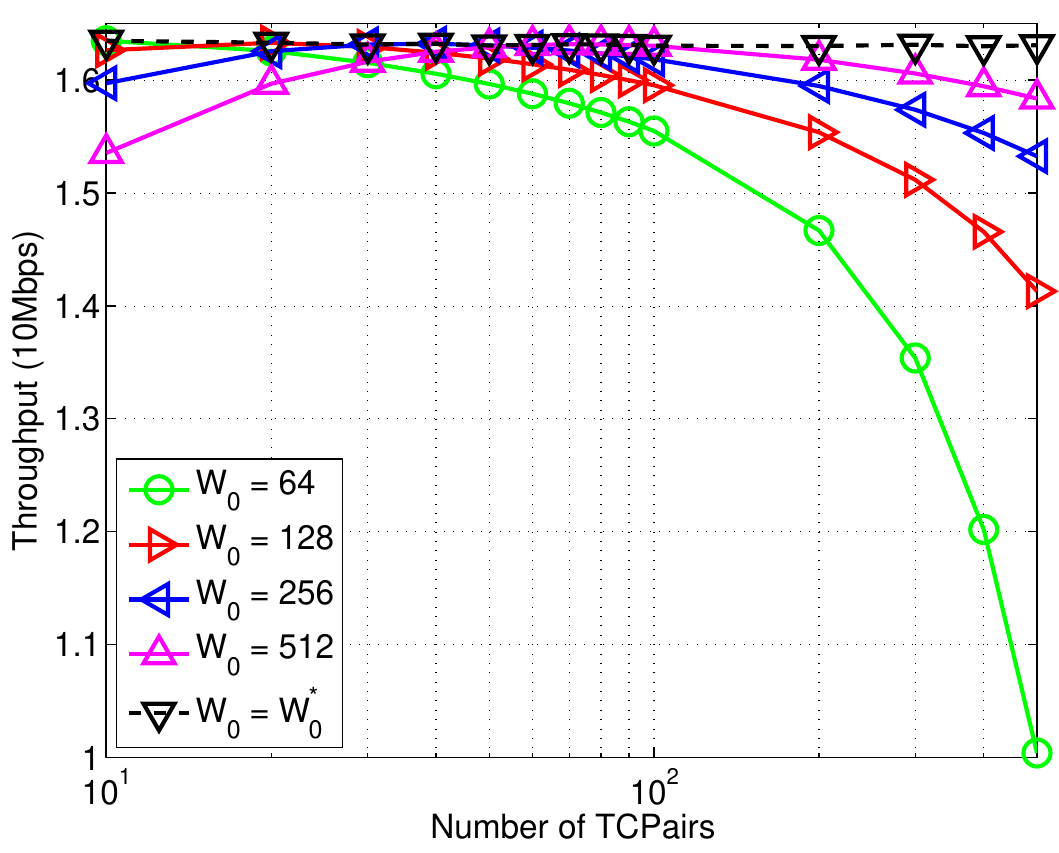}
    \caption{\label{fig:W_opt} The throughput with different $W_0$.}
	\end{minipage}%
	\begin{minipage}[t]{0.49\textwidth}
	\centering
	\includegraphics[width=0.95\textwidth]{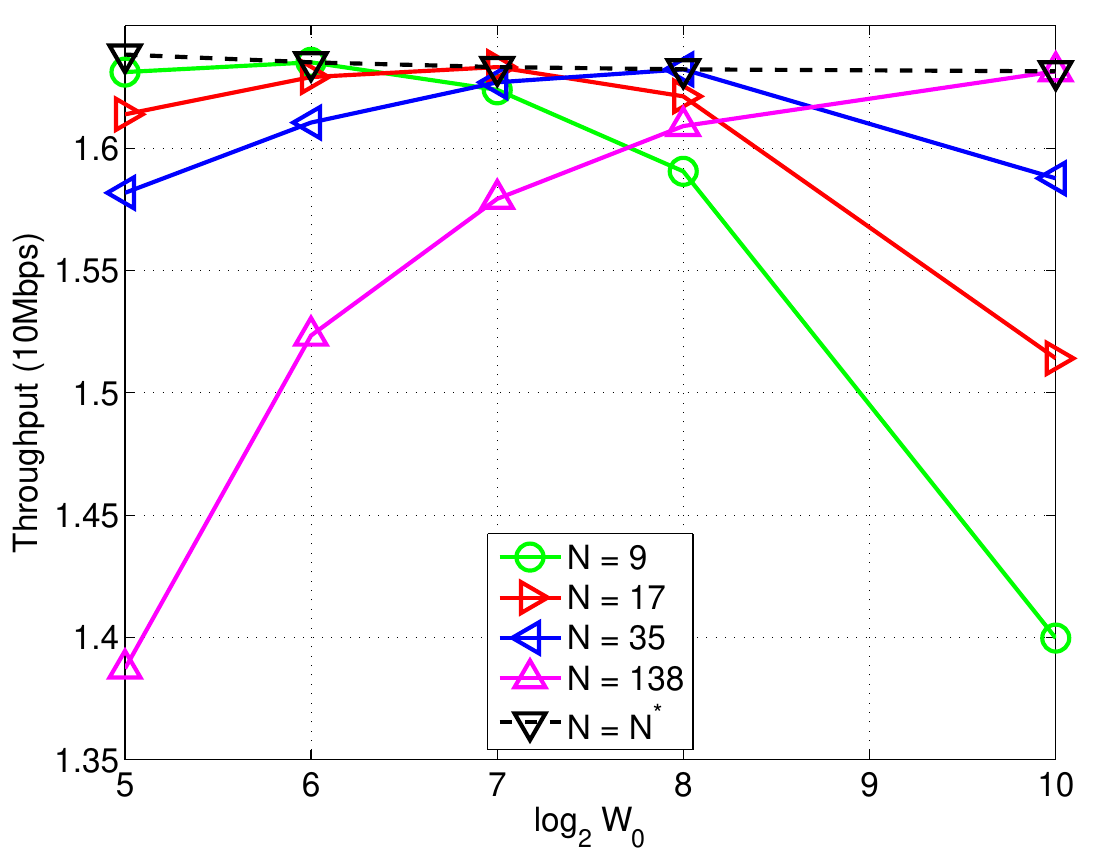}
	\caption{\label{fig:N_opt} The throughput with different $N$.}
	\end{minipage}%
\end{figure}

\begin{table}[tbp]
\caption{Specification of optimal $W_0^*$ and $N^*$.}
\label{tb:optimal_parameters}
\centering
\begin{tabular}{|c|c|c|c|c|c|c|}  % {lccc} ��ʾ����Ԫ�ض��뷽ʽ��left-l,right-r,center-c
\hline
Given $N$ & 20 & 50 & 100 & 200 & 500 \\
\hline
Optimal $W^*_0$ & 128 & 256 & 512 & 1024 & 4096 \\
\hline
Maximum throughput $C$ & 16.33Mbps & 16.31Mbps & 16.30Mbps & 16.30Mbps & 16.30Mbps \\
\hline
Given $W_0$ & 32 & 64 & 128 & 256 & 1024 \\
\hline
Optimal $N^*$ & 4 & 9 & 17 & 35 & 138\\
\hline
Maximum throughput $C$ & 16.38Mbps & 16.35Mbps & 16.33Mbps & 16.32Mbps & 16.31Mbps \\
\hline
\end{tabular}
\end{table}

\begin{figure}
	\begin{minipage}[t]{0.49\textwidth}
	\centering
	\includegraphics[width=0.95\textwidth]{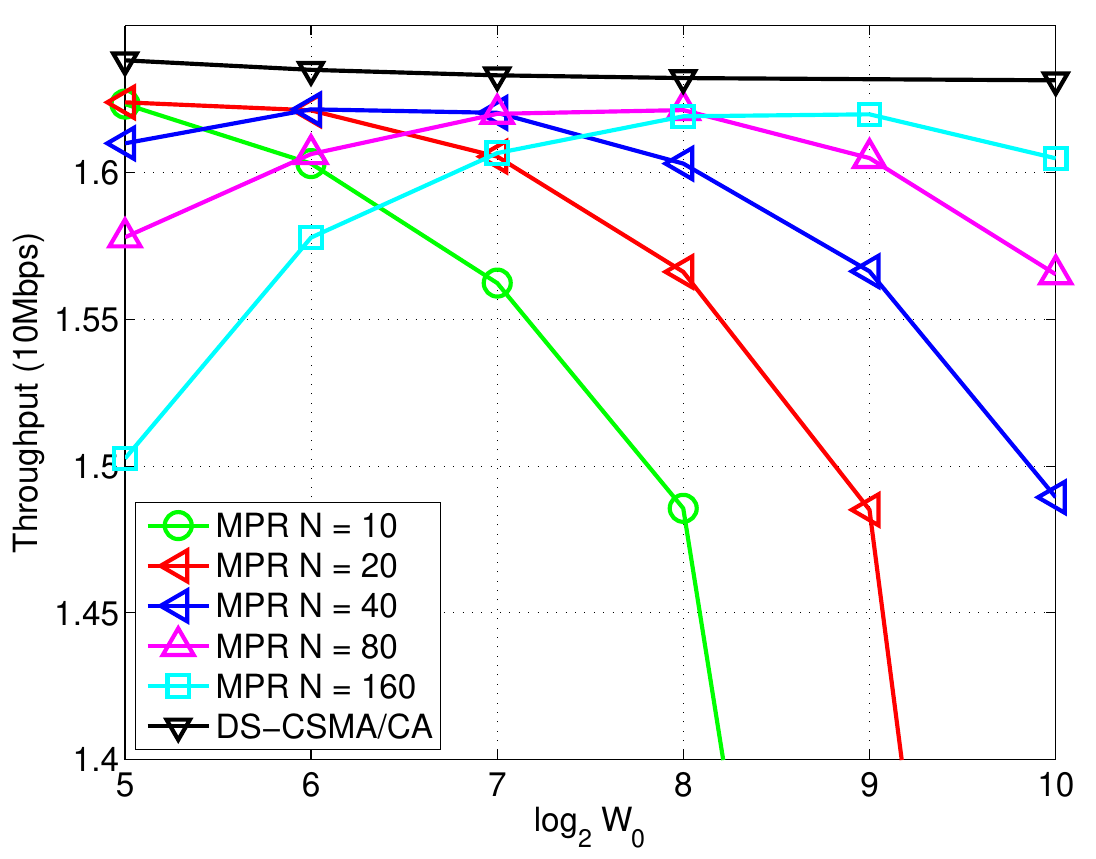}
    \caption{\label{fig:throughput_vs_MPR} The throughput of DS-CSMA compared with MPR.}
	\end{minipage}%
	\begin{minipage}[t]{0.49\textwidth}
	\centering
	\includegraphics[width=0.95\textwidth]{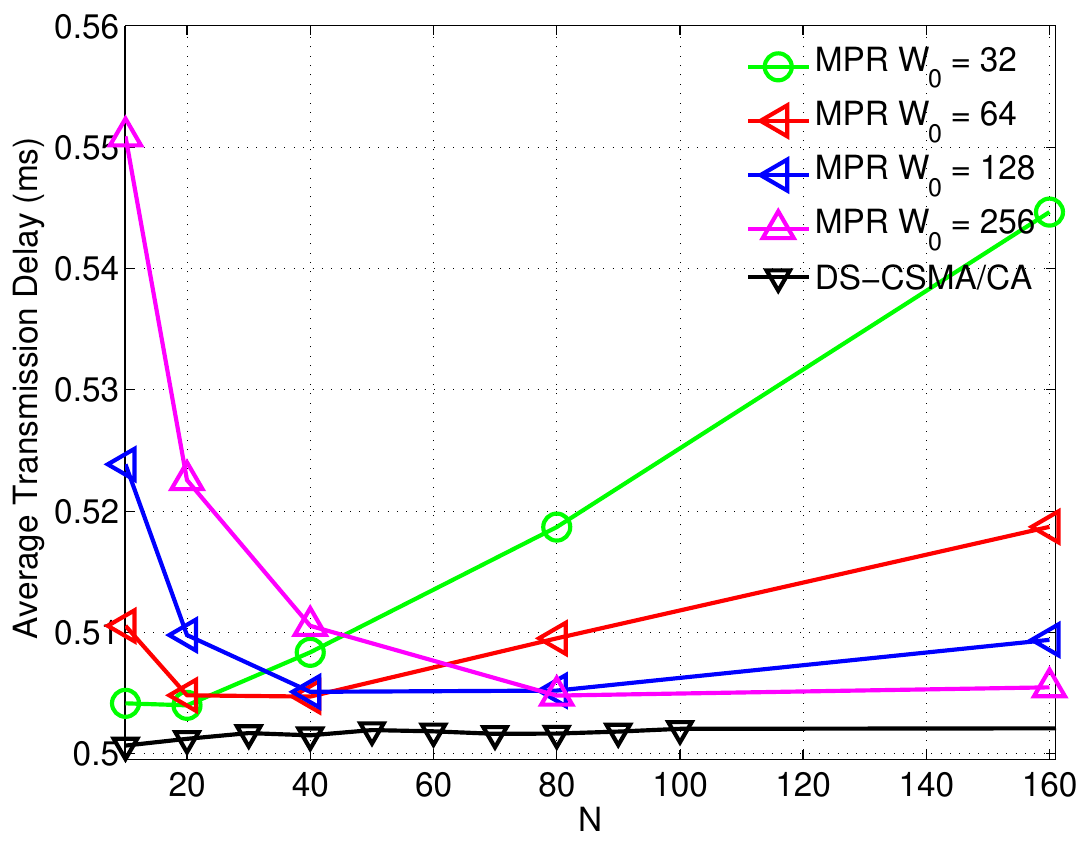}
	\caption{\label{fig:delay_vs_MPR} The average transmission delay of DS-CSMA compared with MPR.}
	\end{minipage}%
\end{figure}

\section{Conclusion}

In this work, based on the physical-layer multi-user communication with symbol boundary misalignment, we have proposed a DS-CSMA protocol for OWSCN, which can avoid collision and enhance the overall throughput.
Furthermore, we have proposed a state transition model for the collision probability and throughput analysis, and for optimizing the initial contention window and indicator matrix.
Both numerical and simulation results show that the proposed DS-CSMA protocol with optimal initial contention window and indicator matrix can significantly achieve higher throughput and lower transmission delay than CSMA/CA including that with MPR.

%\end{spacing}
%
%\begin{spacing}{1.45}

\section{Appendix}

\subsection{Derivation of State Probabilities in Section III}

For case $1$, $m=n=0$, $1 \leq i \leq j \leq W_0-1$,
\begin{equation}
\begin{aligned}
\label{eq:ste_pro_proof1}
\mathbb{P}(0,0,i,j) &= \mathbb{P}(0,0,i+1,j+1) + \sum^{M-1}_{m=0} \sum^{M-1}_{n=0} \sum^{W_m-1}_{i'=0} \mathbb{P}(0,0,i,j | m,n,i',0) \mathbb{P}(m,n,i',0)
\\
&+ \sum^{M-1}_{m=0} \sum^{M-1}_{n=0} \sum^{W_n-1}_{j'=0} \mathbb{P}(0,0,i,j | m,n,0,j') \mathbb{P}(m,n,0,j')
\\
&= \mathbb{P}(0,0,i+1,j+1) + \frac{1-p}{W_0^2} \sum^{M-1}_{m=0} \sum^{M-1}_{n=0} \sum^{W_m-1}_{i'=0} \mathbb{P}(m,n,i',0) + \sum^{W_n-1}_{j'=0} \mathbb{P}(m,n,0,j')
\\
&= \mathbb{P}(0,0,i+1,j+1) + \eta\frac{1-p}{W_0^2}.
%\sum^{W_0-1-j}_{k=1} \mathbb{P}(0,0,i+k,j+k) + \eta\frac{1-p}{W_0^2}
\end{aligned}
\end{equation}

For case $2$, $0<m<M-1,n=0$, $1 \leq i < W_m-1, 0 \leq j \leq W_0-1$,
\begin{equation}
\begin{aligned}
\label{eq:ste_pro_proof1}
\mathbb{P}(m,0,i,j) &= \mathbb{P}(m,0,i+1,j+1) + \mathbb{P}(m,0,i,j | m-1,0,0,j+1) \mathbb{P}(m-1,0,0,j+1)
\\
&= \mathbb{P}(m,0,i+1,j+1) + \frac{\mathbb{P}(m-1,0,0,j+1)}{W_m}p.
\end{aligned}
\end{equation}

For cases $3-7$, the proof of Equations (\ref{eq:ste_Pr3})$-$(\ref{eq:ste_Pr7}) is similar to that of case 2, and it is omitted.

\subsection{Proof of Theorem 1}

For case $1$, $m=n=0$, we have the following equation for $1 \leq i \leq j \leq W_0-1$,
\begin{equation}
\begin{aligned}
\label{eq:ste_updating_proof1}
\mathbb{P}(0,0,i,j) &= \mathbb{P}(0,0,i+1,j+1) + \eta\frac{1-p}{W_0^2}
= \mathbb{P}(0,0,i+2,j+2) + 2\eta\frac{1-p}{W_0^2}
\\
&= (W_0-j) \eta\frac{1-p}{W_0^2}.
%\sum^{W_0-1-j}_{k=1} \mathbb{P}(0,0,i+k,j+k) + \eta\frac{1-p}{W_0^2}
\end{aligned}
\end{equation}
Similarly, we have $\mathbb{P}(0,0,i,j) = \mathbb{P}(0,0,j,i)$ for $1 \leq i,j \leq W_0-1$.
With $i=j=0$, we have $\epsilon_{0,0} = \eta(1-p)W_0^{-1}$.
Furthermore, letting $i=0,1<j\leq W_0-1$, we have $r_{0,0,j} = d_{0,0,j} = (W_0-j) \eta\frac{1-p}{W_0^2} = (W_0-j)W_0^{-1}\epsilon_{0,0}$.

For case $2$, $0<m<M-1,n=0$, we have that
\begin{equation}
\begin{aligned}
\mathbb{P}(m,0,i,j) &= \mathbb{P}(m,0,i+1,j+1) + W_m^{-1}p \mathbb{P}(m-1,0,0,j+1)
\\
&= \mathbb{P}(m,0,i+2,j+2) + W_m^{-1}p [ \mathbb{P}(m-1,0,0,j+2) + \mathbb{P}(m-1,0,0,j+1)]
\\
&= W_m^{-1}p \sum^{W_0-1-j}_{k=1} \mathbb{P}(m-1,0,0,j+k),
\end{aligned}
\end{equation}
when $0\leq i \leq j + W_m - W_0+1 \leq W_m-1$; and
\begin{equation}
\begin{aligned}
\mathbb{P}(m,0,i,j) = W_m^{-1}p \sum^{W_0-1-i}_{k=1} \mathbb{P}(m-1,0,0,j+k),
\end{aligned}
\end{equation}
when $0 \leq j + W_m - W_0+1 < i \leq W_m-1$.

Letting $i=j=0$, we have $\epsilon_{m,0}$ given by
\begin{equation}
\begin{aligned}
\epsilon_{m,0} = \mathbb{P}(m,0,0,0) = W_m^{-1}p \sum^{W_0-1}_{k=1} \mathbb{P}(m-1,0,0,k);
\end{aligned}
\end{equation}
when $i=0$ and $1\leq j \leq W_m - W_0-1$, $r_{m,0,j}$ is given by
\begin{equation}
\begin{aligned}
r_{m,0,j} = \mathbb{P}(m,0,0,j) = W_m^{-1}p \sum^{W_0-1-j}_{k=1} \mathbb{P}(m-1,0,0,j+k);
\end{aligned}
\end{equation}
when $1\leq i \leq W_m - W_0+1$ and $j=0$, $d_{m,0,i}$ is given by
\begin{equation}
\begin{aligned}
d_{m,0,i} = \mathbb{P}(m,0,i,0) = W_m^{-1}p \sum^{W_0-1}_{k=1} \mathbb{P}(m-1,0,0,k);
\end{aligned}
\end{equation}
when $W_m - W_0+1 < i \leq W_m-1$ and $j=0$, $d_{m,0,i}$ is given by
\begin{equation}
\begin{aligned}
d_{m,0,i} = \mathbb{P}(m,0,i,0) = W_m^{-1}p \sum^{W_0-1-i}_{k=1} \mathbb{P}(m-1,0,0,k).
\end{aligned}
\end{equation}
Note that $\boldsymbol{A}_{r,r,m,n}$ and $\boldsymbol{A}_{r,d,m,n}$ are given in Figures \ref{fig:A_rrmn} and \ref{fig:A_rdmn}, respectively, we can obtain the transition equations as shown in Equation (\ref{eq:ste_updating2}).

For case 3, as similar to case 2, we have that
\begin{equation}
\begin{aligned}
\boldsymbol{\mathrm{r}}_{M-1,0} &= W_{M-1}^{-1} p \boldsymbol{\mathrm{A}}_{r,r,m,n} (\boldsymbol{\mathrm{r}}_{M-2, 0} + \boldsymbol{\mathrm{r}}_{M-1, 0}),
\\
\boldsymbol{\mathrm{d}}_{M-1,0} &= W_{M-1}^{-1} p \boldsymbol{\mathrm{A}}_{r,d,m,n} (\boldsymbol{\mathrm{r}}_{M-2, 0} + \boldsymbol{\mathrm{r}}_{M-1, 0}),
\\
\epsilon_{M-1,0} &= W_{M-1}^{-1} p \boldsymbol{1}^T_{W_n-1} (\boldsymbol{\mathrm{r}}_{M-2, 0} + \boldsymbol{\mathrm{r}}_{M-1, 0}).
\end{aligned}
\end{equation}
Consequently, we have the expression of $\boldsymbol{\mathrm{r}}_{M-1,0}$ as follows,
\begin{equation}
\begin{aligned}
\boldsymbol{\mathrm{r}}_{M-1,0} = (\boldsymbol{I} - W_{M-1}^{-1} p \boldsymbol{\mathrm{A}}_{r,r,m,n})^{-1} ( W_{M-1}^{-1} p \boldsymbol{\mathrm{A}}_{r,r,m,n} \boldsymbol{\mathrm{r}}_{M-2, 0} ).
\end{aligned}
\end{equation}

For cases $4-6$, the proof of Equations (\ref{eq:ste_updating4})$-$(\ref{eq:ste_updating6}) is similar to that of cases 2 and 3, and we omit it here.
For case $7$, we can directly reach the conclusion from Equation (\ref{eq:ste_Pr7}).

\subsection{Proof of Theorem 2}

For case $1$, $m=n=0$, according to Equation (\ref{eq:ste_updating_proof1}), we have that $\mathbb{P} (0,0,i,0) = \mathbb{P} (0,0,i,1) = \cdots = \mathbb{P} (0,0,i,i) = \mathbb{P} (0,0,i-1,i) = \cdots = \mathbb{P} (0,0,0,i) = (W_0-i)W_0^{-1}$ for $0\leq i \leq W_0-1$, and
we can hereby separate the items of $i=W_0-1$ and $j=W_0-1$ from $\mathbb{P}(0,0) = \sum_{j=0}^{W_0-1} \sum_{i=0}^{W_0-1} \mathbb{P} (0,0,i,j)$ to obtain the following result,
\begin{equation}
\begin{aligned}
\label{eq:sigma00_proof1}
\mathbb{P}(0,0)
%&= \sum_{j=0}^{W_0-1} \sum_{i=0}^{W_0-1} \mathbb{P} (0,0,i,j)
%\\
&= \mathbb{P} (0,0,W_0-1,W_0-1) + \!\! \sum_{i=0}^{W_0-2} \mathbb{P} (0,0,i,W_0-1) + \!\! \sum_{j=0}^{W_0-2} \mathbb{P} (0,0,W_0-1,j) + \!\! \sum_{j=0}^{W_0-2} \sum_{i=0}^{W_0-2} \mathbb{P} (0,0,i,j)
\\
&= (2W_0-1) W_0^{-1} \epsilon_{0,0} + \sum_{j=0}^{W_0-2} \sum_{i=0}^{W_0-2} \mathbb{P} (0,0,i,j).
\end{aligned}
\end{equation}
Similarly, we can also separate the items of $i=W_0-2$ as well as $j=W_0-2$, and $\mathbb{P}(0,0)$ is simplified as follows,
\begin{equation}
\begin{aligned}
\label{eq:sigma00_proof2}
\mathbb{P}(0,0) &= (2W_0\!-\!1) W_0^{-1} \! \epsilon_{0,0} \!+\! \mathbb{P} (0,\!0,\!W_0\!-\!2,\!W_0\!-\!2) \!+\!\!\! \sum_{i=0}^{W_0-3} \!\! \mathbb{P} (0,\!i,\!W_0\!-\!2) \!+\!\!\! \sum_{j=0}^{W_0-3} \!\!\mathbb{P} (0,\!0,\!W_0\!-\!2,\!j)
\!+\!\!\! \sum_{j=0}^{W_0-3} \! \sum_{i=0}^{W_0-3} \!\! \mathbb{P} (0,\!0,\!i,\!j)
\\
&= (2W_0-1) W_0^{-1} \epsilon_{0,0} + (2W_0-3) 2 W_0^{-1} \epsilon_{0,0} + \sum_{j=0}^{W_0-3} \sum_{i=0}^{W_0-3} \mathbb{P} (0,0,i,j)
\\
&= W_0^{-1} \epsilon_{0,0} \sum_{i=1}^2(2W_0-2i+1)i + \sum_{j=0}^{W_0-3} \sum_{i=0}^{W_0-3} \mathbb{P} (0,0,i,j).
\end{aligned}
\end{equation}
After separating the items of $i,j=0,1,\cdots,W_0-1$, we have that
%corresponding to Equaiton (\ref{eq:sigma_r_d1}),
\begin{equation}
\begin{aligned}
\label{eq:sigma00_proof3}
\mathbb{P}(0,0) &= W_0^{-1} \epsilon_{0,0} \sum_{i=1}^{W_0}(2W_0-2i+1)i
= \frac{1}{6} (2 W_0 + 1) (W_0+1) \epsilon_{0,0} .
\end{aligned}
\end{equation}

For case $2$, $0<m<M-1,n=0$, we have that $\mathbb{P}(m,0) = \sum_{j=0}^{W_0-1} \sum_{i=0}^{W_m-1} \mathbb{P} (m,0,i,j)$.
%\begin{equation}
%\begin{aligned}
%\label{eq:sigma_m0_proof1}
%\mathbb{P}(m,0) = \sum_{j=0}^{W_0-1} \sum_{i=0}^{W_m-1} \mathbb{P} (m,0,i,j).
%\end{aligned}
%\end{equation}
According to Equation (\ref{eq:ste_Pr2}), we have
$\mathbb{P} (m,0,i,0) = \mathbb{P} (m,0,i+1,1) + W_m^{-1} p r_{m-1,0,1}$ for $0\leq i \leq W_m-2$, and
$\mathbb{P} (m,0,W_m-1,0) = W_m^{-1} p r_{m-1,0,1}$.
We separate the items of $j=0$ and combine them with those of $j=1$, which is given as follows,
\begin{equation}
\begin{aligned}
\mathbb{P}(m,0) &= \sum_{i=0}^{W_m-1} \mathbb{P} (m,0,i,0) + \sum_{i=0}^{W_m-1} \mathbb{P} (m,0,i,1) + \sum_{j=2}^{W_0-1} \sum_{i=0}^{W_m-1} \mathbb{P} (m,0,i,j)
\\
&= \sum_{k=1}^2 \sum_{i=0}^{W_m-k} \mathbb{P} (m,0,i,1) + W_m^{-1} p W_m r_{m-1,0,1} + \sum_{j=2}^{W_0-1} \sum_{i=0}^{W_m-1} \mathbb{P} (m,0,i,j).
\end{aligned}
\end{equation}
Similarly, $\mathbb{P} (m,0,i,1) = \mathbb{P} (m,0,i+1,2) + W_m^{-1} p r_{m-1,0,2}$ for $0\leq i \leq W_m-2$ and
$\mathbb{P} (m,0,W_m-1,0) = W_m^{-1} p r_{m-1,0,2}$.
Separating the items of $j=1$ and combining them with those of $j=2$, we have
\begin{equation}
\begin{aligned}
\mathbb{P}(m,0) &= \sum_{i=0}^{W_m-1} \mathbb{P} (m,0,i,2) + \sum_{k=1}^2 \sum_{i=0}^{W_m-k} \mathbb{P} (m,0,i,1) + W_m^{-1} p W_m r_{m-1,0,1} + \sum_{j=3}^{W_0-1} \sum_{i=0}^{W_m-1} \mathbb{P} (m,0,i,j)
\\
&= \sum_{k=1}^3 \sum_{i=0}^{W_m-k} \mathbb{P} (m,0,i,2) + W_m^{-1} p \sum_{j=1}^2 r_{m-1,0,j} \sum_{k=1}^{j} \sum_{i=k}^{W_m} 1 + \sum_{j=3}^{W_0-1} \sum_{i=0}^{W_m-1} \mathbb{P} (m,0,i,j).
\end{aligned}
\end{equation}
Then, we separate the items of $j=2,3,\cdots,W_0-1$, and have the following equation,
\begin{equation}
\begin{aligned}
\mathbb{P}(m,0) &= W_m^{-1} p \sum_{j=1}^{W_0} r_{m-1,0,j} \sum_{k=1}^{j} \sum_{i=k}^{W_m} 1
= W_m^{-1} p \sum_{j=1}^{W_0} r_{m-1,0,j} \bigg[ -\frac{1}{2} j^2 + \bigg(W_m + \frac{1}{2} \bigg) j \bigg].
\end{aligned}
\end{equation}

For cases $3-6$, the demonstration for Equations (\ref{eq:sigma_r_d3})$-$(\ref{eq:sigma_r_d6}) can be achieved similar to case 2, and is omitted here;
for case $7$, we can directly reach the conclusion from Equation (\ref{eq:ste_Pr7}).

\subsection{Proof of Theorem 3}

%In order to calculate $\widetilde{W}_0^* = \arg \max C$,
We obtain $\widetilde{W}_0^*$ by solve the equation $\widetilde{W}_0^* = \arg \{ \eta = \eta^{*} \}$, where $\eta^{*} = \arg \{ \frac{\partial C}{\partial \eta} = 0 \}$.
From Equations (\ref{eq:throughput_expression}) and (\ref{eq:L_o}), we have that
\begin{equation}
\begin{aligned}
\frac{\partial C}{\partial \eta} = \frac{2 L_p \tau}{L_o^2 [T_s + \tau L_o^{-1} - T_c]^2} \frac{\partial L_o}{\partial \eta} \triangleq  C_0 C'_{\eta},
\end{aligned}
\end{equation}
where $C_0 = 2 L_p \tau N^{-1} \eta^{-2} (1-\eta)^{-N} [T_s + \tau L_o^{-1} - T_c]^{-2} > 0$, and
\begin{equation}
\begin{aligned}
C'_{\eta} = (1-\eta)^N - T_c \tau^{-1} \Big\{ N \eta - \big[ 1 - (1-\eta)^N \big] \Big\}.
\end{aligned}
\end{equation}
Hence, $\eta^{*} = \arg \{ \frac{\partial C}{\partial \eta} = 0 \} = \arg \{ C'_{\eta} = 0 \}$.
Based on the assumption that $\eta << 1$, we adopt the conclusion of \cite{bianchi2000performance} to calculate $\eta^{*}$ by
\begin{equation}
\begin{aligned}
\label{eq:eta*}
\eta^{*} = \arg \{ C'_{\eta} = 0 \} \approx \sqrt{2} (N \gamma)^{-1},
\end{aligned}
\end{equation}
where $\tau = \sqrt{T_c \tau^{-1}}$.

We furtherly consider to approximate the proposed optimization to uniform contention window with $W_{\max} = W_0$.
Accordingly, based on Equation (\ref{eq:sigma_r_d2}), we have that $\epsilon_{0,0} \approx \Bigg[ \frac{1}{3} W_0^2 + \frac{1}{2} W_0 + \frac{1}{6} \Bigg]^{-1}$,
%$\epsilon_{0,0}$ given by
%\begin{equation}
%\begin{aligned}
%\epsilon_{0,0} \approx \Bigg[ \frac{1}{3} W_0^2 + \frac{1}{2} W_0 + \frac{1}{6} \Bigg]^{-1},
%\end{aligned}
%\end{equation}
and $\eta$ can be calculated by
\begin{equation}
\begin{aligned}
\label{eq:eta_W0}
\eta &\approx \sum_{i=1}^{W_0}\mathbb{P} (0,0,i,0) + \sum_{j=1}^{W_0}\mathbb{P} (0,0,j,0)
= (W_0 - 1) \Bigg[ \frac{1}{3} W_0^2 + \frac{1}{2} W_0 + \frac{1}{6} \Bigg]^{-1}.
%\\
%&= 2 \epsilon_{0,0} \sum_{i=1}^{W_0} \frac{W_0 - i}{W_0}
%\\
%&= (W_0 - 1) \epsilon_{0,0}
%\\
%&
\end{aligned}
\end{equation}
Then, solving $(W_0 - 1) \big[ \frac{1}{3} W_0^2 + \frac{1}{2} W_0 + \frac{1}{6} \big]^{-1} = \sqrt{2} (N \gamma)^{-1}$, we have $\widetilde{W}_0^*$ as follows,
\begin{equation}
\begin{aligned}
\label{eq:W_0}
\widetilde{W}_0^* \approx \frac{3}{2 \eta^{*}} - \frac{3}{4} + \sqrt{\frac{9}{4 \eta^{* 2}} - \frac{21}{8 \eta^{*}} + \frac{1}{16}}.
\end{aligned}
\end{equation}
Substituting $\eta^*$ in Equation (\ref{eq:eta*}) into Equation (\ref{eq:W_0}), we can readily obtain $\widetilde{W}_0^*$ as Equation (\ref{eq:optimal_W_0}).

\subsection{Proof of Theorem 4}

Based on Equation (\ref{eq:L_o}) and the chain rule of derivation, we have that
\begin{equation}
\begin{aligned}
\label{eq:L_o_N}
\frac{\partial L_o}{\partial N} = \frac{\partial L_o (\eta,N)}{\partial \eta} \frac{\partial \eta}{\partial N} + \frac{\partial L_o (\eta,N)}{\partial N}.
\end{aligned}
\end{equation}

We also consider to approximate the solution of proposed optimization by where $W_{\max} = W_0$.
In this case, from Equation (\ref{eq:eta_W0}), we have $\frac{\partial \eta}{\partial N} = 0$.
Hence, $\widetilde{N}^*$ satisfies
\begin{equation}
\begin{aligned}
\label{eq:L_o_N}
\widetilde{N}^* &=  \arg \bigg\{ \frac{\partial L_o (\eta,N)}{\partial N} = 0 \bigg\}
= \arg \bigg( (1-\eta)^N + T_c \tau^{-1} \Big\{ N \log(1 - \eta) + \big[ 1 - (1 - \eta)^N \big] \Big\} = 0 \bigg).
\end{aligned}
\end{equation}

For $\eta << 1$, we deploy the following approximation for $(1-\eta)^N$ and $\log(1-\eta)$,
\begin{equation}
\begin{aligned}
\label{eq:approx_eta}
(1-\eta)^N \approx 1 - N \eta + \frac{N(N-1)}{2} \eta^2,
\quad
\log(1-\eta) \approx - \eta - \frac{\eta^2}{2}.
\end{aligned}
\end{equation}

Substituting Equation (\ref{eq:approx_eta}) into Equation (\ref{eq:L_o_N}), we have that $\widetilde{N}^*$ approximately satisfies
\begin{equation}
\begin{aligned}
1 - \bigg(\eta + \frac{\eta^2}{2}\bigg) N - \bigg(\frac{T_c}{\tau} - 1\bigg) \frac{\eta^2}{2} N^2 = 0.
\end{aligned}
\end{equation}
%Consequently, $\widetilde{N}^*$ is given by
%\begin{equation}
%\begin{aligned}
%\widetilde{N}^* = \bigg(\frac{T_c}{\tau} - 1\bigg)^{-1} \Bigg[ \sqrt{\frac{\eta^2}{4} + \eta + \frac{2 T_c}{\eta} - 1} - \bigg( 1 + \frac{\eta}{2} \bigg) \Bigg] \eta^{-1},
%\end{aligned}
%\end{equation}
where $\eta$ is given by Equation (\ref{eq:eta_W0}), and the solution is given by Equation (\ref{eq:optimal_N}).

\bibliographystyle{ieeetr}
\bibliography{multistation}

\end{spacing}
\end{document}